\documentclass[twocolumn]{aastex63}

\newcommand{\hii}{H{\sc ii}}

\received{June 1, 2019}
\revised{January 10, 2019}
\accepted{\today}
\submitjournal{ApJ}

\shorttitle{Physical parameters of dust clumps with class I methanol masers}
\shortauthors{Ladeyschikov et al.}

\usepackage{graphicx}
\usepackage{epsfig}
\usepackage{textcomp}
\usepackage{gensymb}
\usepackage{hyperref}
\usepackage{listings}

\begin{document}

\title{The physical parameters of clumps associated with class I methanol masers}

\correspondingauthor{Dmitry A. Ladeyschikov}
\email{dmitry.ladeyschikov@urfu.ru}

\author[0000-0002-3773-7116]{Dmitry A. Ladeyschikov}
\affiliation{Ural Federal University \\ 51 Lenin Str., 620051 Ekaterinburg, Russia}

\author[0000-0002-1605-8050]{{James S. Urquhart}}
\affiliation{Centre for Astrophysics and Planetary Science, University of Kent, Canterbury CT2 7NH, UK}

\author[0000-0002-3773-7116]{Andrey M. Sobolev}
\affiliation{Ural Federal University, 51 Lenin Str., 620051 Ekaterinburg, Russia}

\author[0000-0002-4047-0002]{Shari L. Breen}
\affiliation{SKA Organisation, Jodrell Bank Observatory, SK11 9DL, UK}

\author[0000-0003-4116-4426	]{Olga S. Bayandina}
\affiliation{Joint Institute for VLBI ERIC Oude Hoogeveensedijk 4, 7991 PD Dwingeloo, The Netherlands}

\begin{abstract}

We present a study of the association between class I methanol masers and cold dust clumps from the ATLASGAL survey. It was found that almost 100\% of class I methanol masers are associated with objects listed in the ATLASGAL compact source catalog. We find a statistically significant difference in the flux density, luminosity, number and column density and temperature distributions of ATLASGAL sources associated with 95/44 GHz methanol masers compared with those ATLASGAL sources devoid of 95 GHz methanol masers. The masers tend to arise in clumps with higher densities, luminosities and temperatures compared with both the full sample of the ATLASGAL clumps, as well as the sample of ATLASGAL sources that were cross-matched with positions previously searched for methanol masers but with no detections.  Comparison between the peak position of ATLASGAL clumps and the interferometric positions of the associated class I and II methanol masers reveals that class I masers are generally located at larger physical distances from the peak submillimetre emission than class II masers. We conclude that the tight association between ATLASGAL sources and class I methanol masers may be used as a link toward understanding the conditions of the pumping of these masers and evolutionary stages at which they appear.  

\end{abstract}


\section{Introduction} \label{sec:intro}

Methanol masers are ubiquitous features of shock waves and star formation regions in both our own and other galaxies. On the basis of the works of \cite{Batrla1987} and \citet{Menten1991} methanol masers are divided into two classes -- I and II. Class II methanol masers (MMII) have a radiative{-radiative (source-sink notation)} pumping mechanism \citep{Cragg2005} and are usually found close to the sources of strong radiation. MMII are associated with high-mass star formation regions (SFR) \citep[e.g][]{BIL19}. Class I methanol masers (MMI) have a collisional-radiative pumping mechanism \citep{Sobolev2007} and are usually found at some distance from a radiation source -- in the shock waves that produce suitable conditions to excite these masers. Unlike MMII, MMI trace not only high-mass SFR but also low-mass SFR \citep{Kalenskii2013} and other sites with shock waves, including supernova remnants \citep{Pihlstrom2014}, molecular cloud collisions \citep{Salii2002}, and \hii{} regions interacting with molecular gas \citep{Voronkov2010b}. Association with the shocked regions implies that MMI are produced in the regions where densities and temperatures are elevated. This is in good agreement with results of \deleted{the} model calculations \citep{SOB18,LEU18,VOR06}.

Submillimeter continuum emission directly probes the dense interstellar material from which stars form and therefore where MMI may appear. The ATLASGAL 870 $\mu$m survey \citep{Schuller09}  produced a large-scale, systematic database of massive pre- and proto-stellar clumps in the Galaxy. This survey reveals the location of the highest density regions in the interstellar medium. Recently \citet{Ladeyschikov2019} compiled a database of all the known MMI, providing an opportunity for comparison between the location of these masers with the dust continuum emission at 870 $\mu$m.

Comparison between class II methanol masers and ATLASGAL 870 $\mu$m emission was previously made in \citet{Urquhart13, Urquhart15} and \citet{BIL19}. In each of these papers an association was found between 99\% of MMII detected in the Methanol MultiBeam (MMB) survey \citep{CAS10,GRE10,CAS11,GRE12,BRE15}, with compact, dense clumps from the ATLASGAL catalog \citep{CON13,URQ14}.  The MMB  source sample was matched to ATLASGAL in the range of $280\degr<l<20\degr$ in \citet{Urquhart13}.  In \citet{Urquhart15} a dedicated programme of follow-up APEX observations of $\sim70$ MMB sources was presented. Data from the final part of the MMB catalogue \citep[20\textdegree$<l<$60\textdegree,][]{BRE15}, which was not available on the time of publication of \citet{Urquhart13, Urquhart15}, was analysed in \citet{BIL19}, where the authors identified the host clumps for 958 class II methanol masers across the Galactic Plane and studied their physical parameters  using a combination of ATLASGAL and the JCMT Plane Survey catalogues (JPS; \citealt{moore2015, eden2017}).

The relationship between class I methanol masers and cold dust clumps was previously studied by \citet{CHE12}. The authors conducted a 95~GHz class I methanol maser survey toward color-selected GLIMPSE sources associated with Bolocam 1.1 mm cold dust clumps (\citealt{aguirre2011,dunham2011}). It was reported that clumps associated with class I methanol masers had higher values of column density and integrated flux density. 

In this paper, we compare the full sample of known class I  methanol masers from the maser database \citep{Ladeyschikov2019} with the data from the ATLASGAL survey of cold dust clumps at 870~\micron{}. We used the following transitions of class I methanol masers for matching with the ATLASGAL survey: 95~GHz, 84~GHz, 44~GHz and 36~GHz. We used the observations from more than 100 papers, but the following papers make the most significant contribution to the statistics: \citet{YAN17,YAN20,KIM18,KIM19,CHE11,CHE12,GAN13,BAE11,BRE19,JOR15,JOR17,VAL00}.  The full list of papers incorporated into  the class I maser database and used for analysis  in this paper is available online\footnote{\href{http://maserdb.net/list.pl}{http://maserdb.net/list.pl}}. 

Cross-matching between MMI and dust clumps in the Central Molecular Zone (CMZ) was not considered in this paper, as the masers in this region (such as those detected by \citealt{COT16}) are much more crowded and numerous compared to other parts of the Galactic plane. Other methods and techniques need to be applied to study the association between MMI and dust clumps in this region; the detailed analysis of this region will be presented in a future publication.

\section{Data Processing}
\label{sect:data}
\subsection{Source sample: MMI and cold dust emission}
The class I maser database \citep{Ladeyschikov2019} was used to  combine  the Galactic distribution of all published class I methanol maser observations (single-dish or interferometric), into maser sites. The details of the grouping process for these methanol masers are presented in Section\,2.6 of \citet{Ladeyschikov2019}. 
{In some instances maser sites have been observed on more than one occasion using different facilities. In our analysis we adopt flux densities measured with single-dish telescopes and  where available we use positions derived from interferometric  observations to determine the angular separation between MMI and ATLASGAL sources. In case where there is more than one  positional measurement for  a particular maser  is available, the median value is used. In case where more than one measurement of the flux density is available, the maximum value is used.}  The subsequent analysis in the paper was performed on maser sites, not on the individual maser observations.

The ATLASGAL 870 $\mu$m survey \citep{Schuller09}, obtained using the APEX telescope, is a continuum survey covering the whole inner Galactic plane (280\textdegree$<l<$60\textdegree, $|b|<$1.5\textdegree). The ATLASGAL catalogue \citep{CON13,URQ14} consists of 10163 sources, including 517 that are located in the Central Molecular Zone (359.3\textdegree$<l<$1.7\textdegree, $|b|<$0.2\textdegree). 
This catalogue was produced using the source extraction algorithm \texttt{SExtractor} (\citealt{Bertin1996}) and is 99\% complete at $\sim$6\,$\sigma$, which corresponds to a 870\,$\mu$m peak flux density of 0.3-0.4\,Jy beam$^{-1}$, and  has a positional accuracy of $\sim$4\,arcsec \citep{CON13,URQ14}. The physical properties (distance, dust temperature, luminosity, mass and gas column density) have been determined for approximately 8000 of these dense clumps \citep{Urquhart18}. Matching  the methanol maser samples with these dense clumps  provides a straightforward and convenient way of investigating the physical conditions from which these masers arise. 

The maser database presented in \citet{Ladeyschikov2019} contains 532  maser sits located in the  ATLASGAL region studied in this paper (280\textdegree$<l<$60\textdegree, $|b|<$1.5\textdegree)  excluding $\sim2000$ of 36 GHz maser located in the CMZ \citep{COT16}. The number of masers detected within the region at 95\,GHz is 420, at 44\,GHz is 355, at 84\,GHz is 83, and at 36\,GHz is 95. There is some overlap between objects observed at different frequencies: the number of objects detected at both 95 and 44 GHz is 253,  and 69 sources are detected at all of the frequencies considered.

\subsection{Catalog cross-matching}

\begin{figure}
\fig{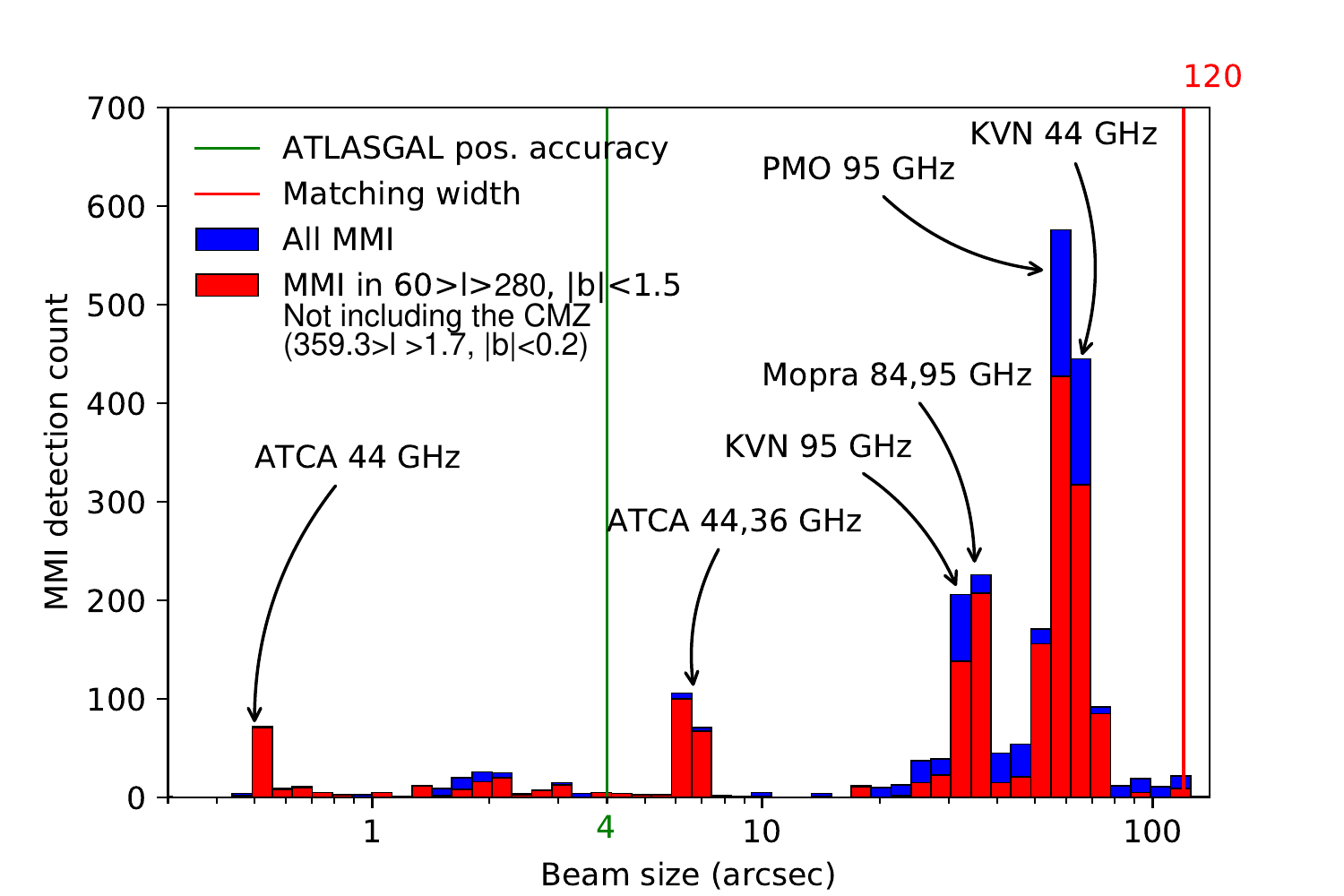}{0.49\textwidth}{}
\caption{Log-based histogram of primary beam sizes of class I methanol maser observations in the studied sample. Blue bars refer to the whole MMI database. Red bars show masers in the currently studied sample (280\textdegree$<l<$60\textdegree, $|b|<1.5\degr$).  The green vertical line indicates ATLASGAL positional accuracy (4 arcsec). The red vertical line shows the matching threshold for the association between MMI and ATLASGAL clumps. Labels mark the facilities that make the most significant contributions to the histogram.}
\label{fig_beam}
\end{figure}

The cross-matching between MMI and ATLASGAL catalogs was done in two independent ways. 

Firstly we looked for matches between the maser sites and ATLASGAL clumps by determining the positional offsets in arcsecs from their catalog positions, and comparing this to the matching radius;  we consider masers and dust clumps to be physically associated if the angular offset is smaller than the matched radius, while  pairs with larger offsets are discarded. If more than one ATLASGAL clump is matched to a particular maser, then the clump with the smallest angular offset to the MMI center is considered to be the most likely association.  

The matching radius depends mainly on the maximum beam size of maser observations and this can vary significantly due to the large range of beam sizes of the studies used to produce the maser database (see Figure~\ref{fig_beam} for telescope beam sizes).
There are $\sim440$ MMI detections with high accuracy  positions (synthesized beam size 0.4-7 arcsec) derived from the VLA, ATCA, ALMA, and SMA interferometers. However, the number of single-dish observations with detections ($\sim$2000; 82\% of all detections) dominates over the number of available interferometric observations. The beam size of single-dish detections varies from 10 arcsec (IRAM 30-m at 229~GHz), 55 arcsec (Mopra 22-m at 95~GHz) to 105 arcsec (Onsala 20-m at 36~GHz). An analysis of beam size distribution  of the combined sample (see Figure~\ref{fig_beam})  suggests that a matching radius of 60 arcsec, corresponding to beam FWHM of 120 arcsec, covers almost all maser observations.  Thus we use a matching radii of 60 arcsec in all subsequent analysis. 

In addition to the source catalogue, SExtractor also produced image masks marking the location and extent of each extracted clump. In these masks the flux values for each clump have been replaced with an integer value that link the pixels  associated with each clump to their catalogue entry. Converting the maser position to the corresponding pixel position on the ATLASGAL masks therefore provides an unambiguous match to the dense clump it is embedded in. We have used these masks as a second independent method to match the ATLASGAL clumps with the{ir embedded} masers. 

This method eliminates the possibility of a false match caused by an irregular morphology or elongation of the dust emission. However, there is still a possibility for false associations in crowded regions. If many ATLASGAL sources are present near the maser position within the matching radius, then we pick the nearest ATLASGAL clump. Given that most ATLASGAL sources are about 60-80 arcsec in size, we can consider that most matches except those made with Onsala Telescope are reliable to make the correct association between dust clumps and masers. the accuracy of the maser position is essential when considering the cross-match between masers and ATLASGAL sources. The list of masers in the current study uses the interferometric maser positions when possible. Otherwise, the median average of single-dish coordinates is used. That eliminates the significant inaccuracies of maser positions in each maser site. However, when we consider the offsets between dust clumps and maser positions, only interferometric maser positions are used.  

We examined the difference in the results of matching obtained using these two methods  and this reveals that $\sim$6\% of the ATLASGAL counterparts differ.  We have visually checked these sources and find that they are mainly localised to crowded regions of ATLASGAL emission where is is difficult to reliably match sources even visually.  Given the small number of sources this applies to, we concluded that these sources do not dominate the general statistics. In further analysis, we exclude these sources as being less reliable.  

\section{Results}

\subsection{Matching between ATLASGAL and MMI sources}\label{Sec_match}

\begin{figure}
\fig{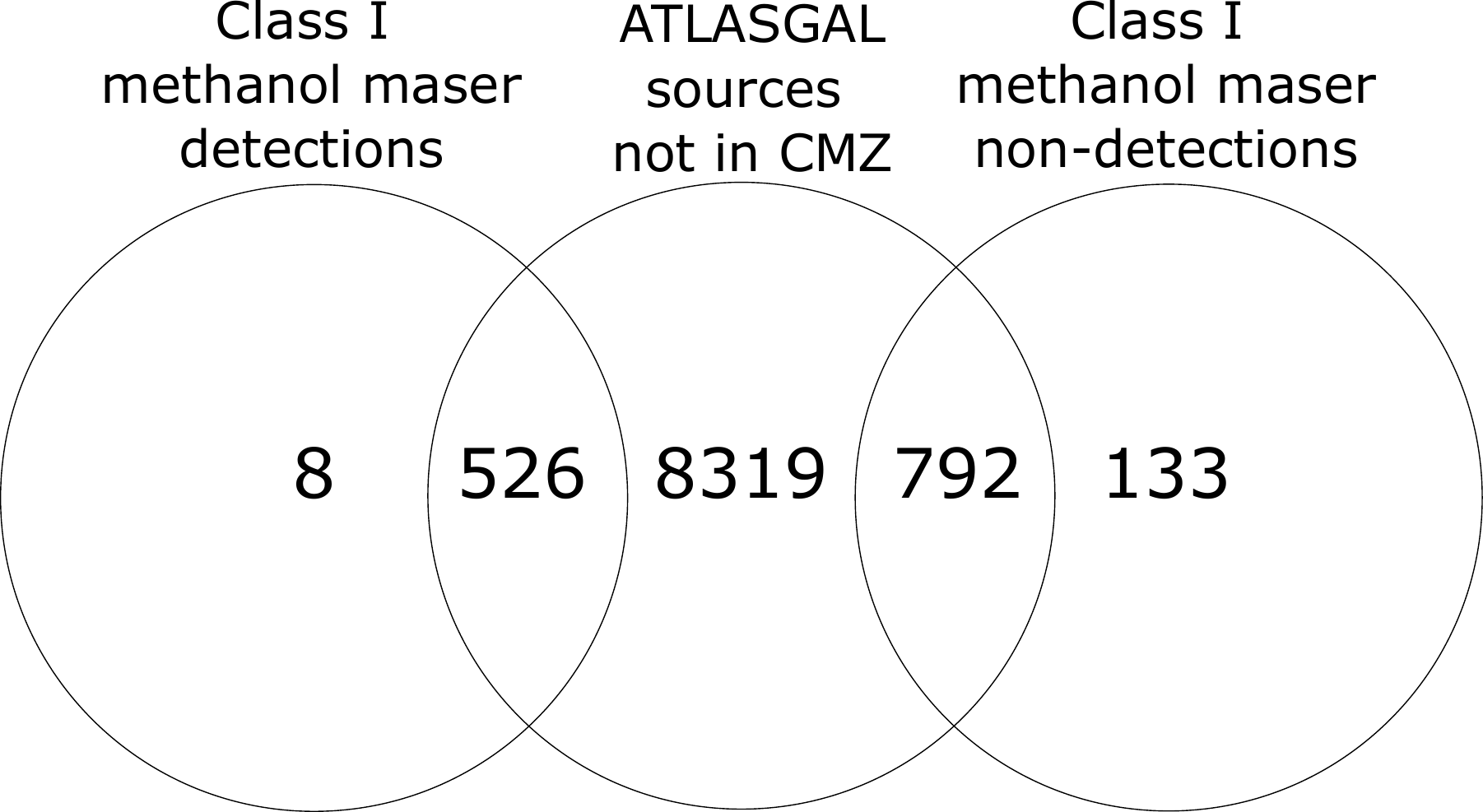}{0.45\textwidth}{}
\caption{Venn diagram presenting the matching statistics for objects with class I methanol masers detections/non-detctions and ATLASGAL counterparts in the region 280\textdegree$<l<$60\textdegree, $|b|<1.5\degr$, excluding the CMZ region. 
\label{fig_venn}}
\end{figure}

From the cross-matching between ATLASGAL sources and class I methanol masers, we have found that almost 100\% class I methanol maser sources are associated with ATLASGAL sources reported in the compact source catalog \citep{CON13,URQ14}. The details of the matching statistics are presented in Figure~\ref{fig_venn}. Given that there are $\sim9600$ ATLASGAL clumps in the region (excluding the CMZ), the percentage of ATLASGAL sources with known class\,I counterparts is $\sim5$\%.

{The maser database also stores the MMI non-detections, thus we can study their association with ATLASGAL clumps. From cross-matching using a 60 arcsec radius, we found that a substantial fraction of objects with a non-detection of MMI (85\% of 923 sources) have an ATLASGAL counterpart. Most of them were observed at 95~GHz. That suggests a detection rate towards ATLASGAL sources of $\sim$30\%.}

\subsection{Completeness of the sample of the detected MMI} \label{sect_complet}

To verify the completeness of the sample of the detected class I methanol maser sources, we match the data of the blind MMI survey MALT-45 \citep{JOR15} with ATLASGAL data using a 60 arcsec association radius. This survey has a Galactic coverage of $330\degr<l<335\degr$, $|b|<0.51\degr$ and 5$\sigma$ sensitivity limit of 4.5 Jy.  Out of 77 detected MMI, 73 are found to be associated with ATLASGAL clumps, leading to the association rate of 95\%.

There are 674 ATLASGAL sources in the region $330\degr<l<335\degr$, $|b|<0.51\degr$ and so the MMI association rate with ATLASGAL clumps is 9.8\%. Given that the overall match rate is $\sim5.5$\% (532 masers in 9646 ATLASGAL sources excluding the CMZ); this might suggest that the \citet{Ladeyschikov2019} MMI catalogue only contains $\sim$56\% of the total population of masers. This makes a strong case for a new blind survey for MMI or a programme of targeted observations towards ATLASGAL sources.

\subsection{MMI without ATLASGAL counterparts} \label{MMI_without_AGAL}

\begin{figure*}
\gridline{
\fig{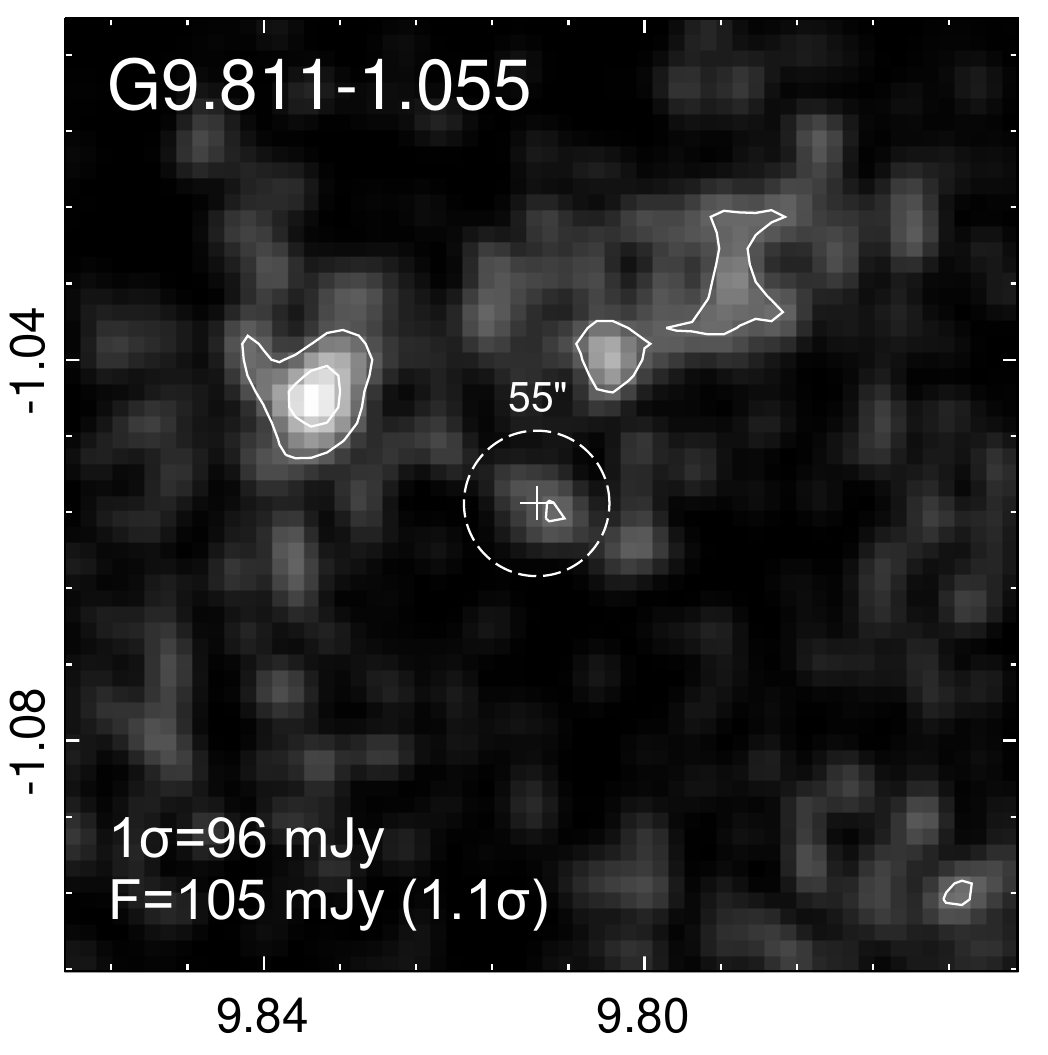}{0.25\textwidth}{(1) G9.811-1.055}
\fig{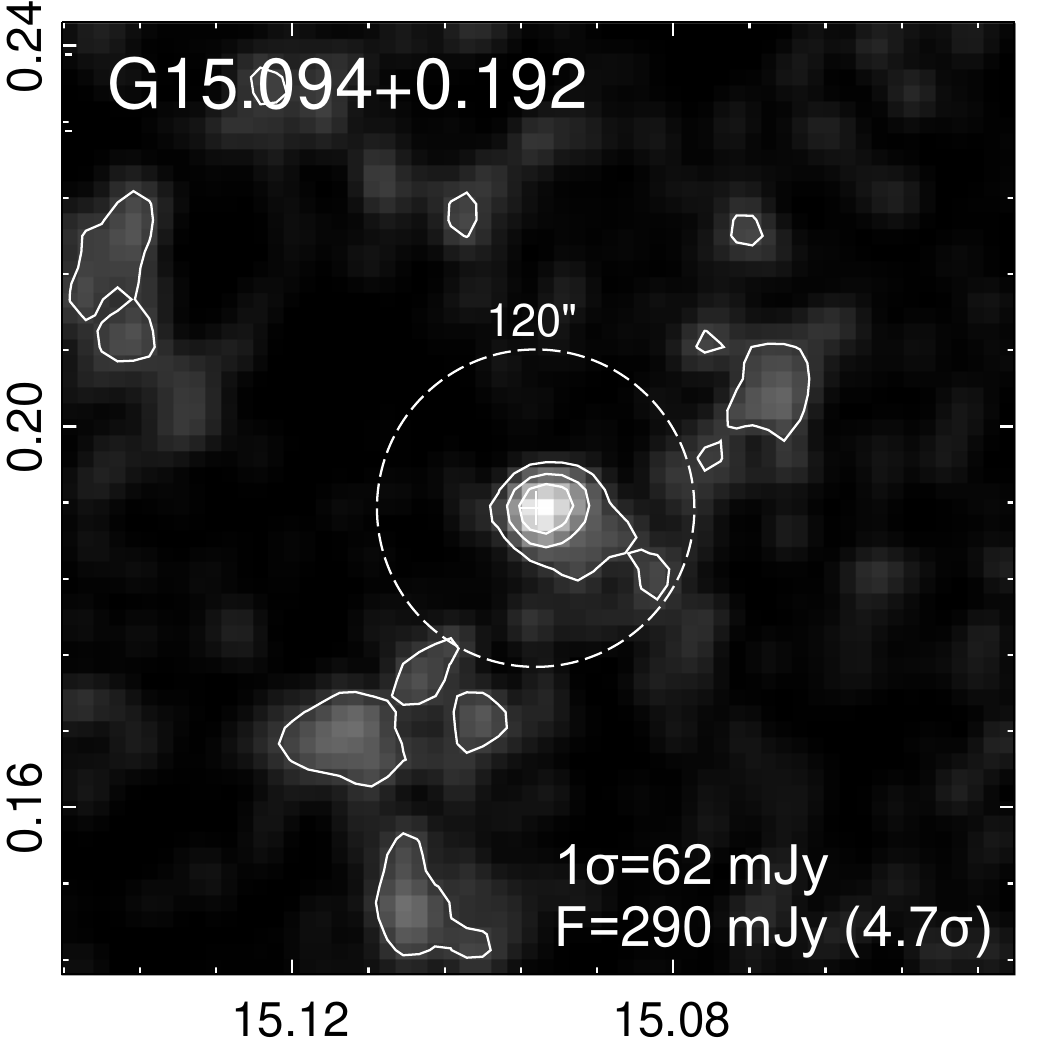}{0.25\textwidth}{(2) G15.094+0.192}
\fig{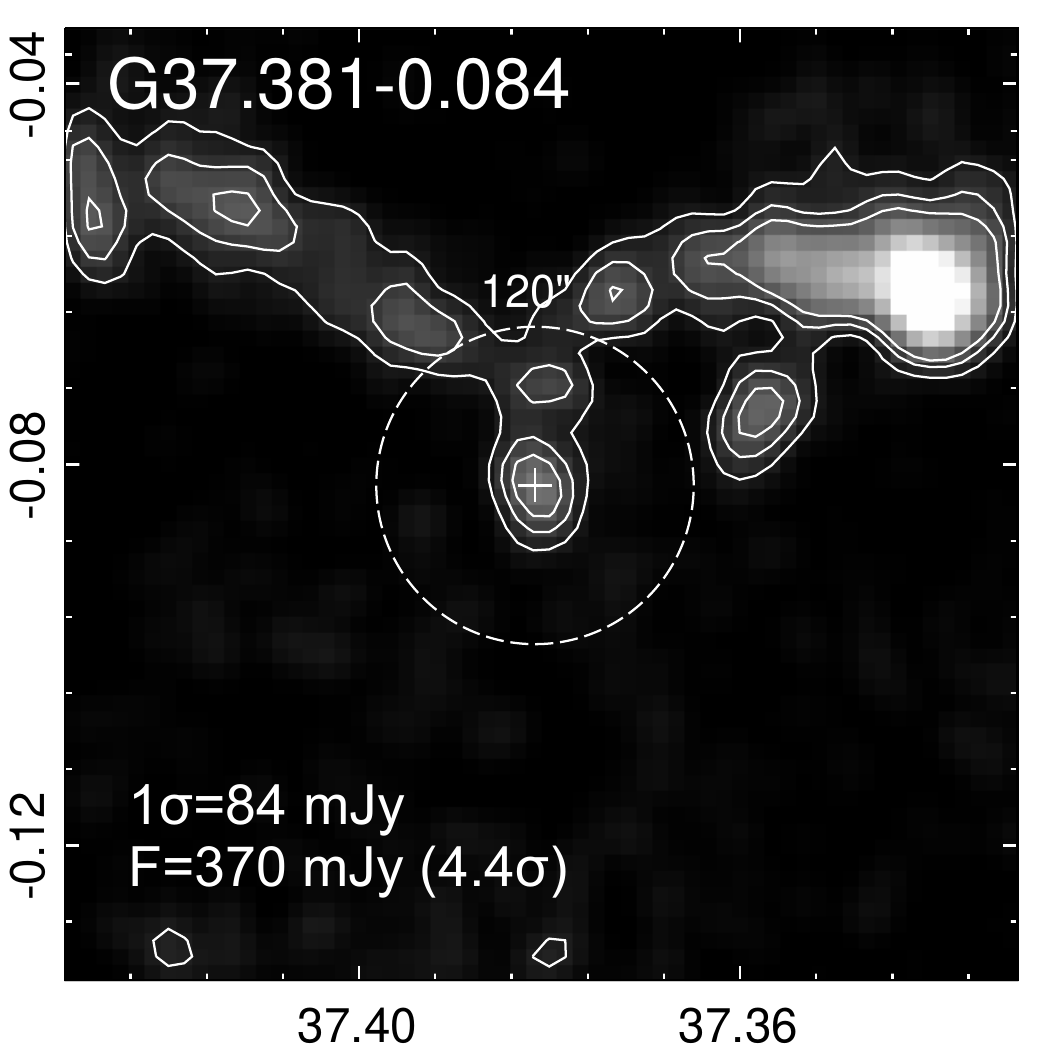}{0.25\textwidth}{(3) G37.381-0.084}
\fig{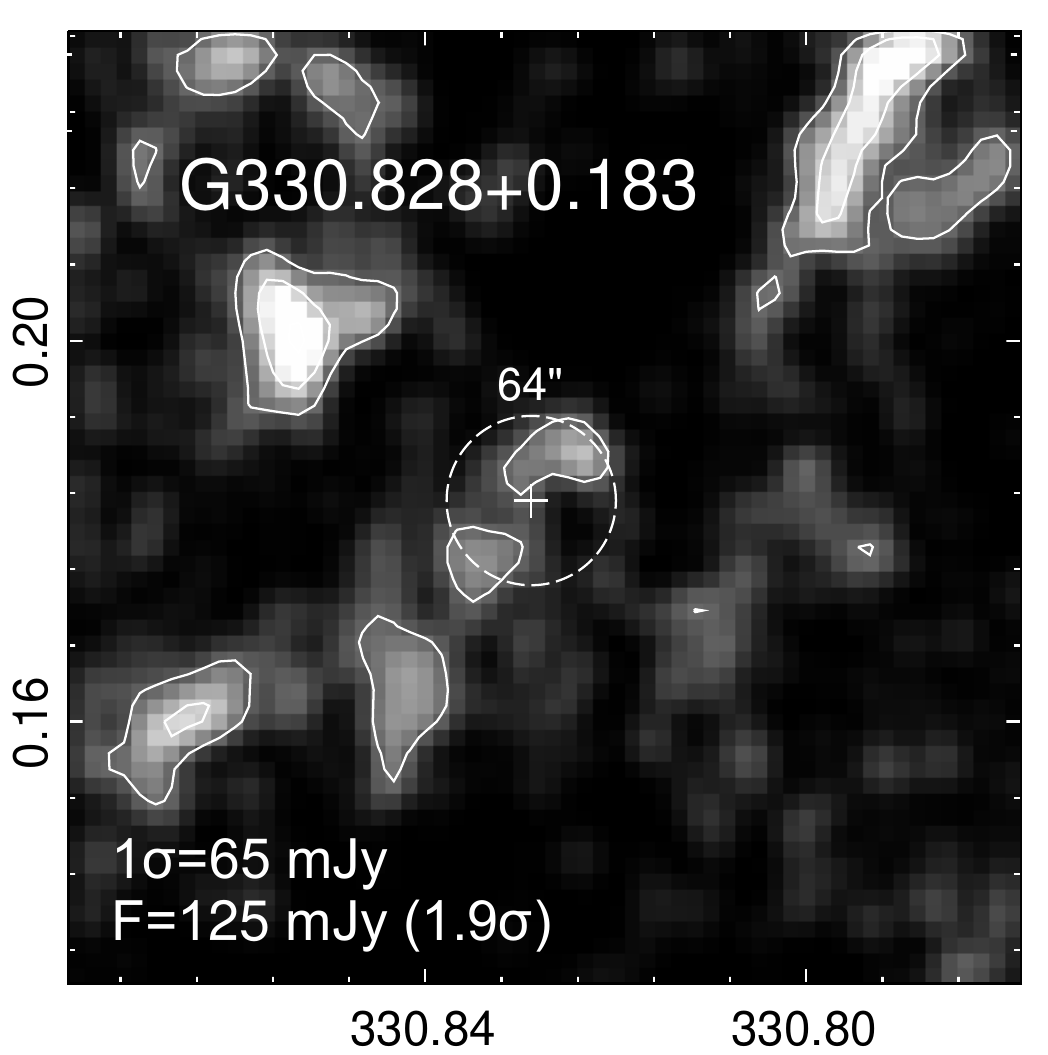}{0.25\textwidth}{(4) G330.828+0.183}
 }
 \gridline{
\fig{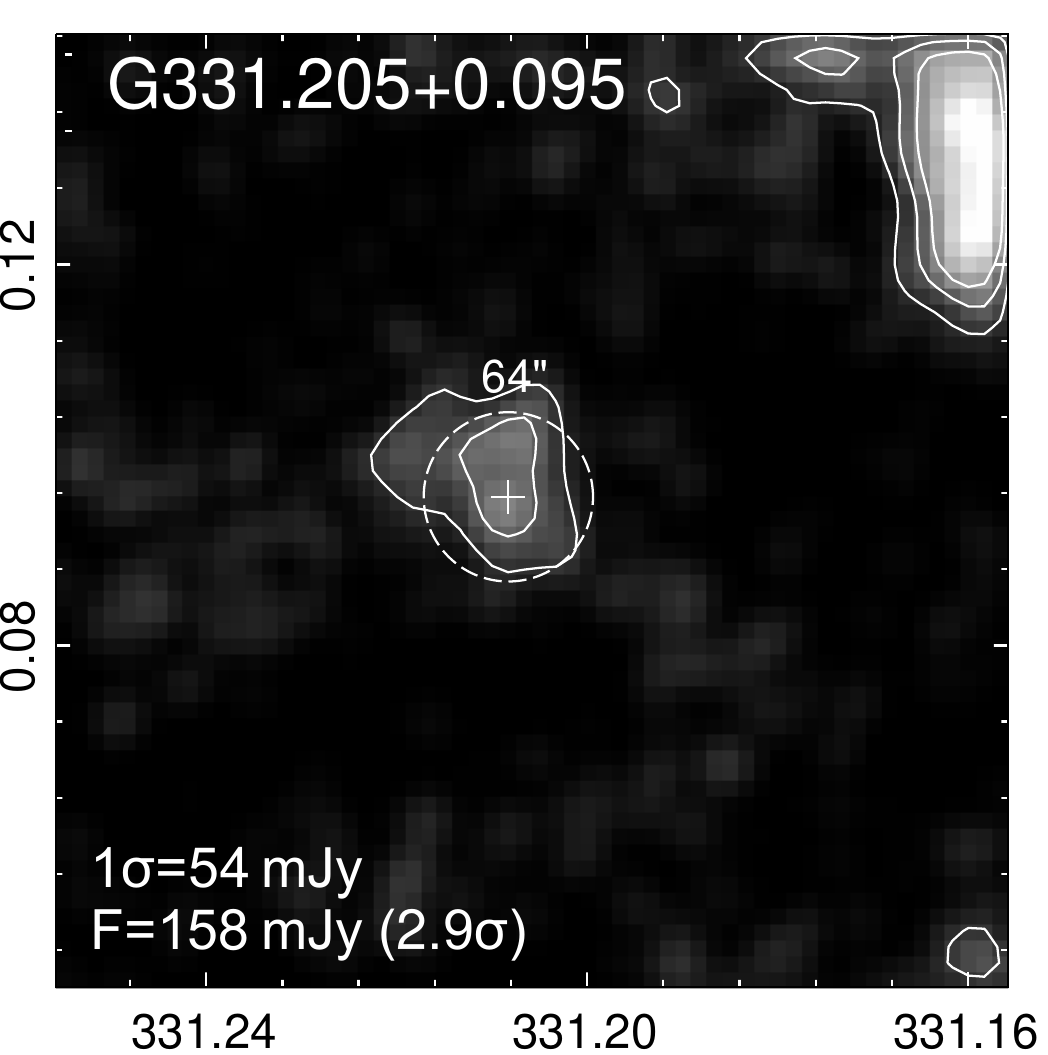}{0.25\textwidth}{(5) G331.205+0.095}
\fig{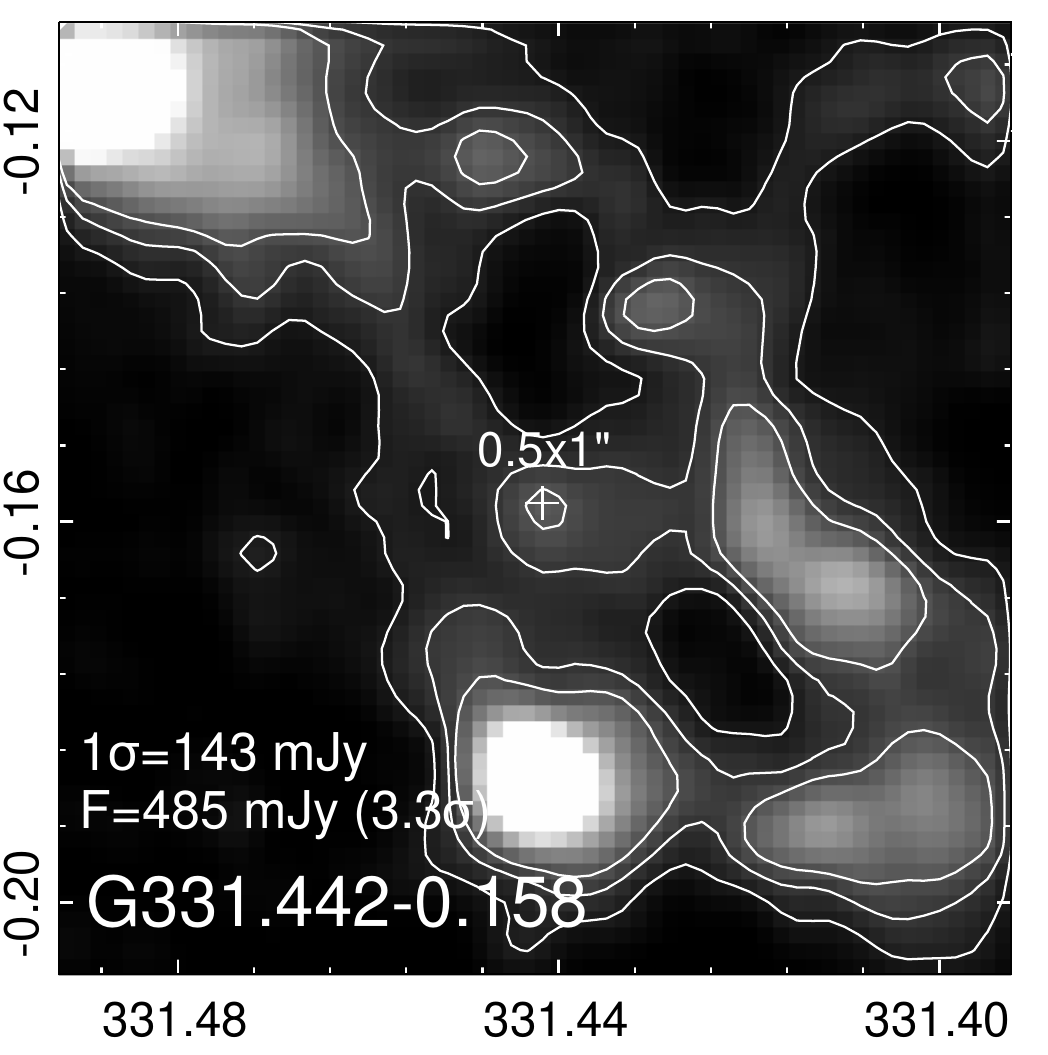}{0.25\textwidth}{(6) G331.442-0.158}
\fig{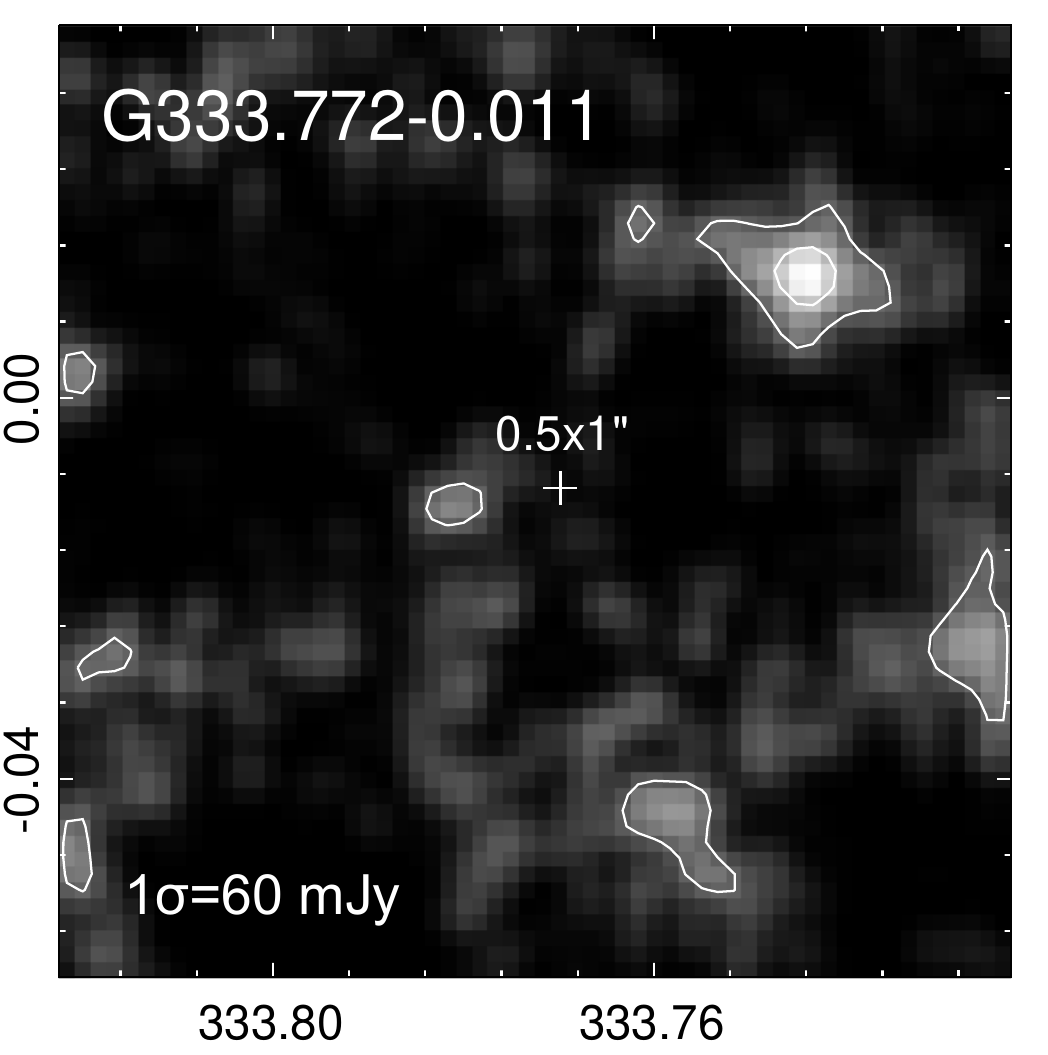}{0.25\textwidth}{(7) G333.772-0.011}
\fig{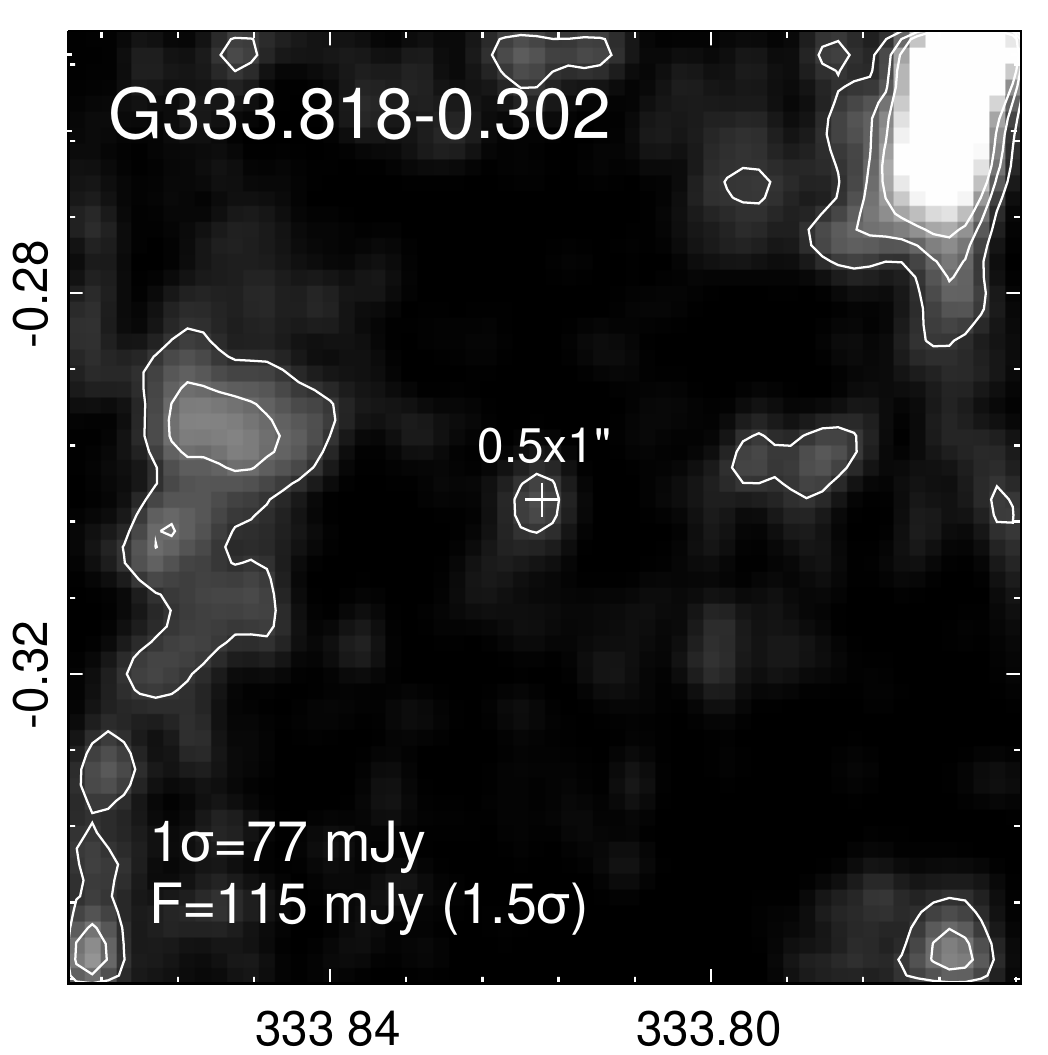}{0.25\textwidth}{(8) G333.818-0.302}
 }
 \caption{Class I methanol maser detections without an ATLASGAL compact source counterpart. The coordinate system in each panel is galactic. The background image in each panel is the ATLASGAL 870 $\mu$m emission. The crosses mark the positions of detected class I methanol maser and the circles represent the beam size of the corresponding observations. Contour levels are 1$\sigma$, 2$\sigma$ and 3$\sigma$ level of 870 $\mu$m emission. In each panel 1$\sigma$ level is described together with 870 $\mu$m flux density at  the maser position.}
   \label{fig_nodust}
\end{figure*}

\begin{deluxetable*}{cccccccccc}
\tabletypesize{\footnotesize}
\tablewidth{0pt}
 \tablecaption{List of MMI without associated ATLASGAL compact sources. \label{tab1}}
 \tablecolumns{10}
 \tablehead{
 \colhead{ID} & \colhead{Source name} & \colhead{RA} & \colhead{Dec}  & \colhead{$V_\mathrm{peak}$(MMI)} & \colhead{$F_\mathrm{peak}$(MMI)} & \colhead{$F_\mathrm{peak}$(870\micron{})}  & \colhead{References$^{1}$} & \colhead{Notes} \\ 
  \colhead{} & \colhead{} & \colhead{(h m s)} & \colhead{(h m s)}  & \colhead{(km~s$^{-1}$)} & \colhead{(Jy)} & \colhead{(mJy)}  & \colhead{} & \colhead{}
 }
 \startdata 
1 & G9.811-1.055 & 18 11 19.1 & -20 57 56 &     32.47-34.48 & 0.93-2.88 & 105 (96)  & YAN17; YAN20 &  \\ 
2 & G15.094+0.192 & 18 17 20.9 & -15 43 46 &30.36 & 3.5 & 290 (62) & KIM18 &  870 $\micron{}$ emission at 4.7$\sigma$   \\ 
3 & G37.381-0.084 & 18 59 51.6 & +03 55 18 & 56 & 3.2 & 370 (84) & KAN15 & 870 $\micron{}$ emission at  4.4$\sigma$  \\ 
4 & G330.828+0.183 & 16 07 41.1 & -51 43 48 &     -82 & 3.1 & 125 (65) & JOR15 & Spurious? \\ 
5 & G331.205+0.095 & 16 09 52.2 & -51 32 19 &     -67.4 & 3.5 & 158 (54) & JOR15 & Spurious? \\ 
6 & G331.442-0.158 & 16 12 04.8 & -51 33 56 &    -85.98 & 0.73 & 485 (143) & JOR17 &  870 $\micron{}$ emission at 3.3$\sigma$ \\ 
7 & G333.772-0.011 & 16 22 00.5 & -49 50 21 &    -89.33;-89.50 & 16 & $<$ 60 & JOR15;JOR17 &  \\ 
8 & G333.818-0.302 & 16 23 29.5 & -50 00 41 &    -47.70;-48.50 & 19 & 115 (77) & JOR15;JOR17 &  870 $\micron{}$ emission at 1.5$\sigma$ \\ 
 \enddata
\tablecomments{References of class I methanol maser detections are as follows: YAN17 -- \citet{YAN17}, YAN20 -- \citet{YAN20} KIM18 -- \citet{KIM18}, JOR15 -- \citet{JOR15}, KAN15 -- \citet{KAN15}, JOR17 -- \citet{JOR17}. \\
}
\end{deluxetable*}

Using the results of the radius-based matching described in Section~\ref{Sec_match} we find associations between 524 MMI and ATLASGAL compact sources within a 60 arcsec match radius. Given that there are 532 known MMI within the ATLASGAL survey range, this leaves 8 MMI ($\sim$1.5\%) devoid of an associated dust continuum source. Details of these masers are presented in Table~\ref{tab1} and Figure~\ref{fig_nodust}. 

The sources G331.442-0.158 and G333.818-0.302 are associated with 870 \micron{} emission within 12 arcsec. However, the ATLASGAL flux density at the maser position of G333.818-0.302 is only 115 mJy, while $\sigma=77$~mJy, leading to \added{a} detection at the 1.5$\sigma$ \added{level}. Maser  G331.442-0.158 appears to be associated with low-level (2-3 $\sigma$) diffuse 870 \micron{} emission, which is difficult to separate into sources. Thus these sources \replaced{wasn't included to}{were not included in} the compact source catalog. Two other sources, G15.094+0.192 and G37.381-0.084, have a weak 870 \micron{} compact emission that is below the $3\sigma$ detection limit of ATLASGAL, thus they also were not included in the compact source catalog.

The remaining five masers do not have an associated ATLASGAL clump and bright extended emission within a 60 arcsec radius.

We show all  eight sources in Figure~\ref{fig_nodust}. Two of the maser sources, G330.828+0.183 and G331.205+0.095, reported by \cite{JOR15} were not detected in follow up observations conducted by \cite{JOR17} and so the maser emission may be either spurious detections or variable sources. Thus, only sources G333.772-0.011 and G9.811-1.055 may be considered as detections of MMI without cold dust emission.  From a multi-wavelength analysis (WISE all-sky \citep{Wright10}, Spitzer IRAC \citep{IRAC}, Hershel PACS \citep{PACS} and SPIRE \citep{SPIRE}, and APEX ATLASGAL) of these sources, it was found that source G333.772-0.011 with a maser flux density of 16 Jy \citep{JOR15,JOR17} is the only example of MMI without a counterpart in the sub-millimeter range. G9.811-1.055, with a MMI flux density of 2.88 Jy \citep{YAN17}, has a counterpart in the Herschel SPIRE 500 \micron{} map and an associated ``green'' source in Spitzer IRAC map. The origin of these masers without ATLASGAL counterparts is unknown and needs further investigation. 

\section{Discussion}

Class I methanol masers are associated with the compression and heating of the gas, presumably created by shock waves. The tight correlation between MMI and ATLASGAL sources can be attributed to their association with star formation regions that provide the necessary conditions ({e.g. heat, density and shocks}) to produce class I methanol masers. Taking into account the analysis of methanol masers without \added{an} ATLASGAL counterpart (see Section~\ref{MMI_without_AGAL}), \replaced{we suggest that association between ATLASGAL dust clumps and class I methanol masers is close to 100\%}{we suggest that in close to 100\% of the cases class I methanol masers are associated with ATLASGAL dust clumps}. However, the  detection of dust emission  does not guarantee the  conditions necessary for masers will be present. In the current paper, we focus only on masers in \deleted{the} star-formation regions and leave the class I methanol masers in other types of objects for  a future  study. The other types of objects include extragalactic class I methanol masers \citep[e.g.][]{Chen16}, masers in the CMZ \citep{COT16}, masers in supernova remnants \citep{McEwen16,Pihlstrom14} and masers in cloud-cloud collisions \citep{Salii02}.

\subsection{Physical parameters of the ATLASGAL clumps associated with MMI}\label{sec:phys_par}

\begin{table}
\caption{{Peak values from the Gaussian fit of the clumps physical parameters histograms in three samples: ATLASGAL clumps associated with masers at 95 GHz; ATLASGAL clumps with no class I maser detection at 95 GHz; the full sample of ATLASGAL clumps with available physical parameters. The estimation of the parameter errors are shown after the $\pm$ symbol. Values in brackets are the standard deviation of the Gaussian fit.}\label{tab_fitpar}}
\setlength{\tabcolsep}{1pt}
\small
\centering
 \begin{tabular}{lcccc}
\hline\noalign{\smallskip}
  Sample &  With maser & No maser &  All clumps  \\
   Size &  442 & 703 &  8002  \\
  \hline
  $\log F_\mathrm{peak}$ & 0.372 $\pm$ 0.013 & 0.008 $\pm$ 0.008 & -0.245 $\pm$ 0.007  \\ 
  & $<0.11>$ &	$<0.07>$ &	$<0.08>$ \\ 
 $\log N_\mathrm{H_2}$ & 22.784 $\pm$ 0.013 & 22.442 $\pm$ 0.009 & 22.28 $\pm$ 0.005 \\
  & $<0.12>$ &	$<0.08>$ &	$<0.04>$ \\  
  $\log L_\mathrm{bol}$ & 3.905 $\pm$   0.038 &    3.382 $\pm$ 0.035 &	2.868 $\pm$ 0.018 \\
  & $<0.07>$ &	$<0.08>$ &	$<0.03>$ \\
  $\log T_\mathrm{dust}$ & 1.339 $\pm$  0.005 &    1.313 $\pm$ 0.003 &	1.268 $\pm$ 0.002 \\
  & $<0.9>$ &	$<0.07>$ &	$<0.05>$ \\  
  $\log M_\mathrm{FWHM}$ & 2.714 $\pm$ 0.033 &	   2.657 $\pm$ 0.028 &	2.587 $\pm$ 0.013 \\
  & $<0.04>$ &	$<0.04>$ &	$<0.01>$ \\  
  $\log  n\mathrm{(H_2)}$ & 4.958 $\pm$ 0.016 & 4.519 $\pm$ 0.016 & 4.564 $\pm$ 0.008  \\
  & $<0.09>$ &	$<0.06>$ &	$<0.03>$ \\  
  $\log  L_\mathrm{bol}/M_\mathrm{FWHM}$ & 1.281 $\pm$ 0.911 & 0.023 $\pm$ 0.022 & 0.65 $\pm$ 0.013  \\
  & $<0.09>$ &	$<0.09>$ &	$<0.05>$ \\

 \hline
  {\smallskip}
\end{tabular}
\end{table}

We utilize the catalog of ATLASGAL physical clump properties \citep{Urquhart18} to study the relationship between the physical parameters of clumps and the detection of class I methanol masers. For clump radii and mass, we recalculate the values using 870 \micron{} emission above the FWHM flux contour, similar to \citet{BIL19}. Thus the clump mass is named FWHM clump mass in further analysis. This was done to eliminate the effect of radii overestimation in evolved clumps. As the embedded source evolve and heat their environment, more of the \replaced{clumps}{clump} outer envelop becomes detectable, thus more evolved clumps tend to be bigger. However, this is an observational bias and not a real  evolutionary trend \citep{Urquhart18}. 

\begin{figure*}
\gridline{\fig{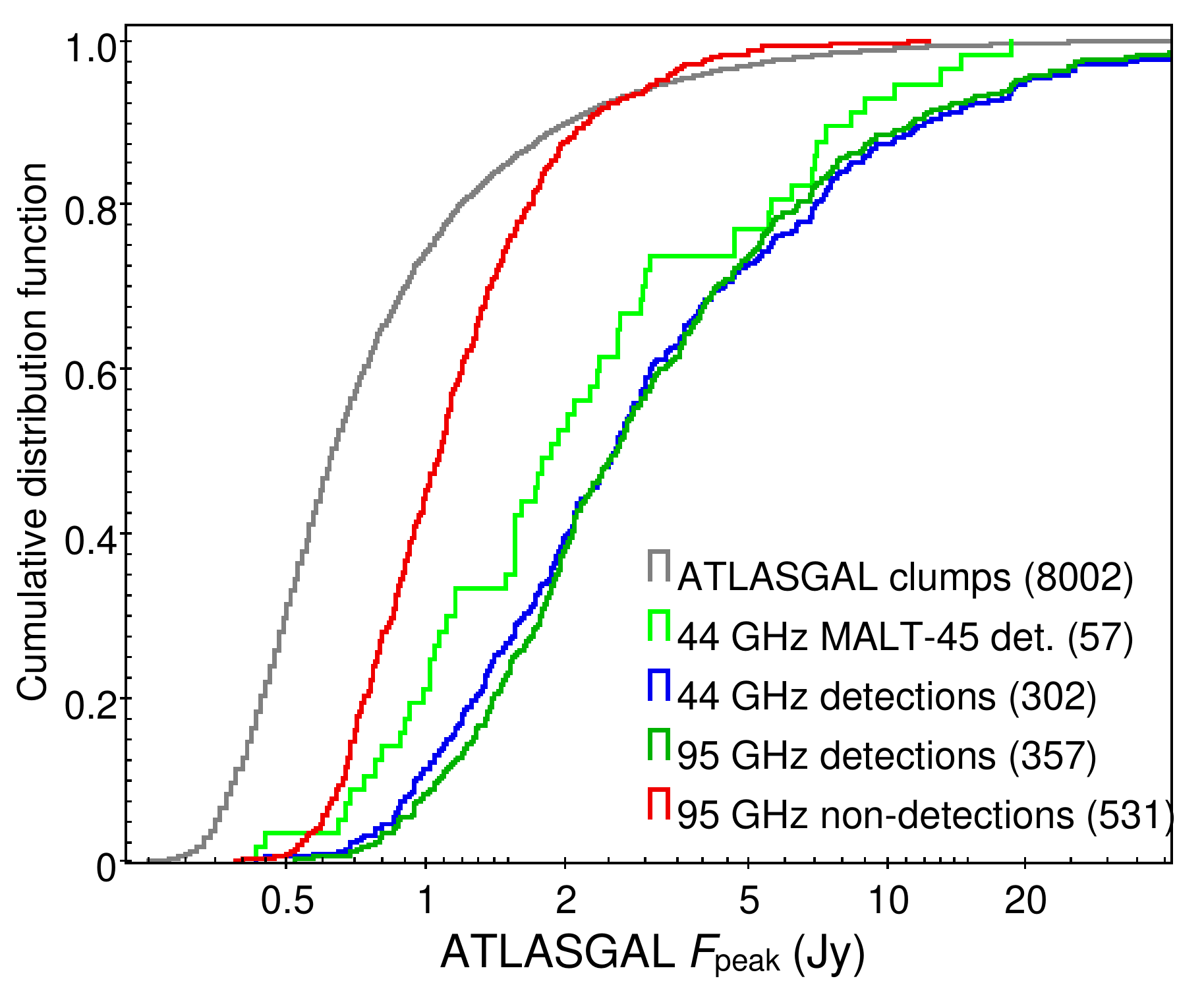}{0.40\textwidth}{(A)}
\fig{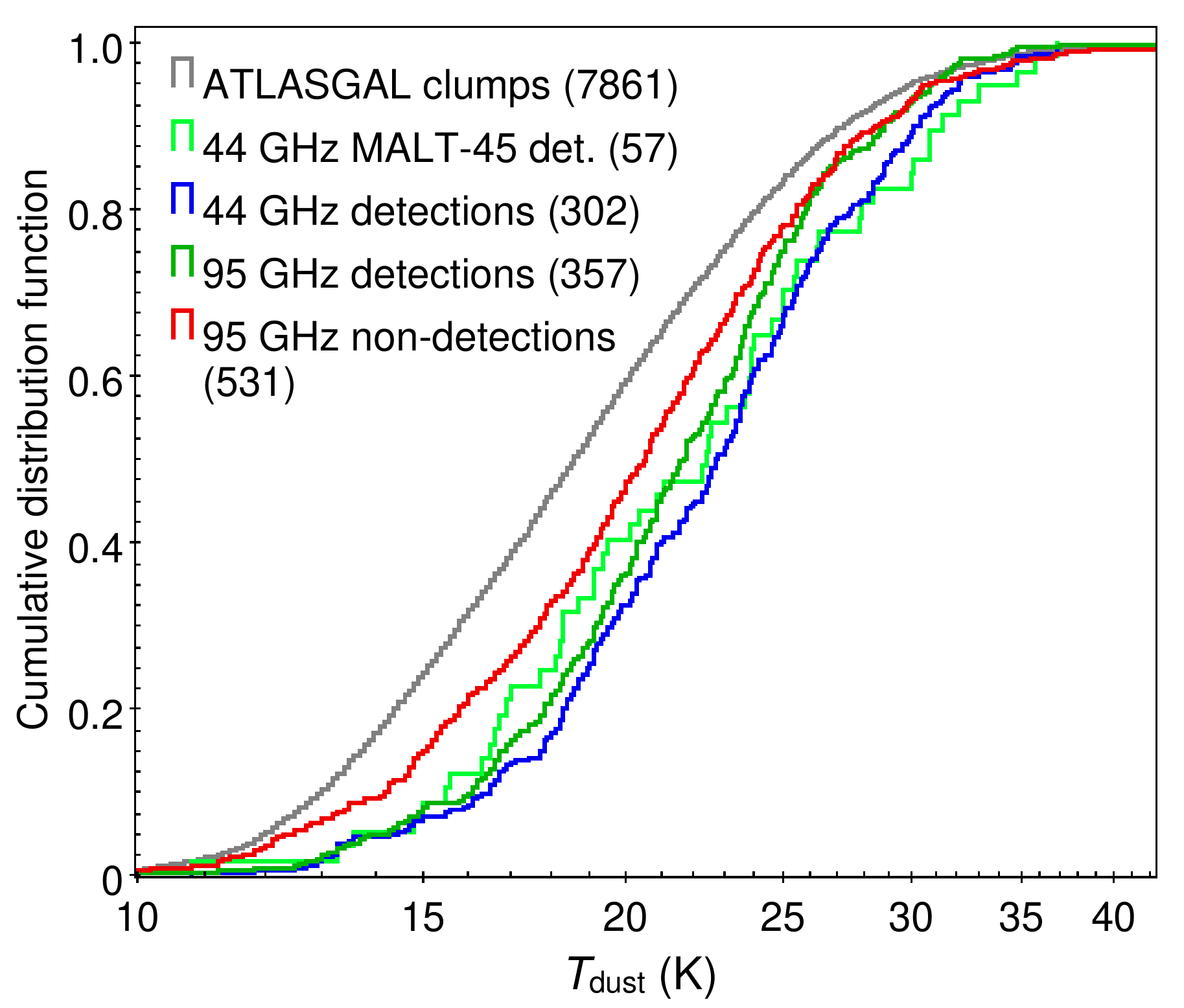}{0.40\textwidth}{(B)}}
\gridline{\fig{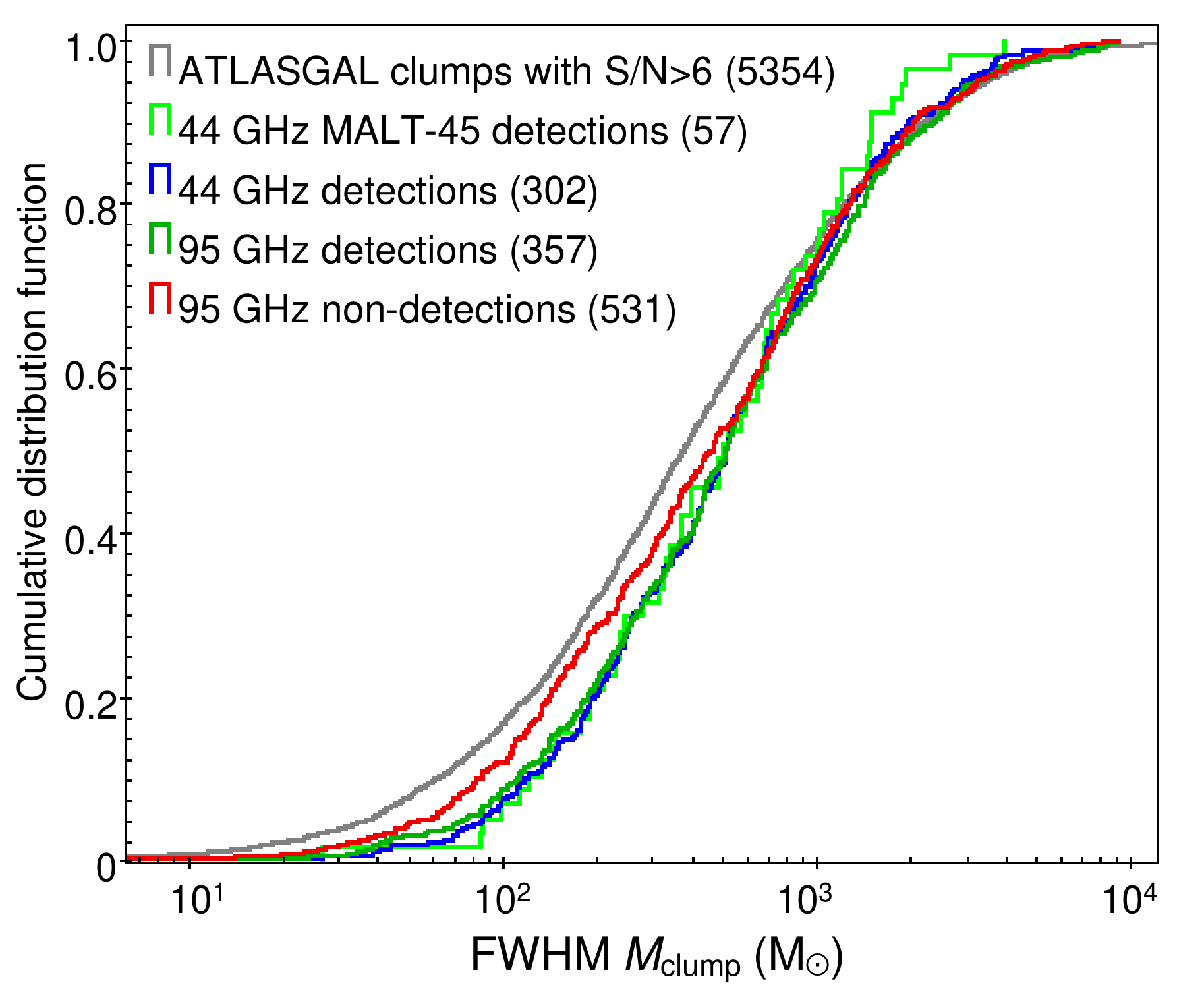}{0.40\textwidth}{(C)}
\fig{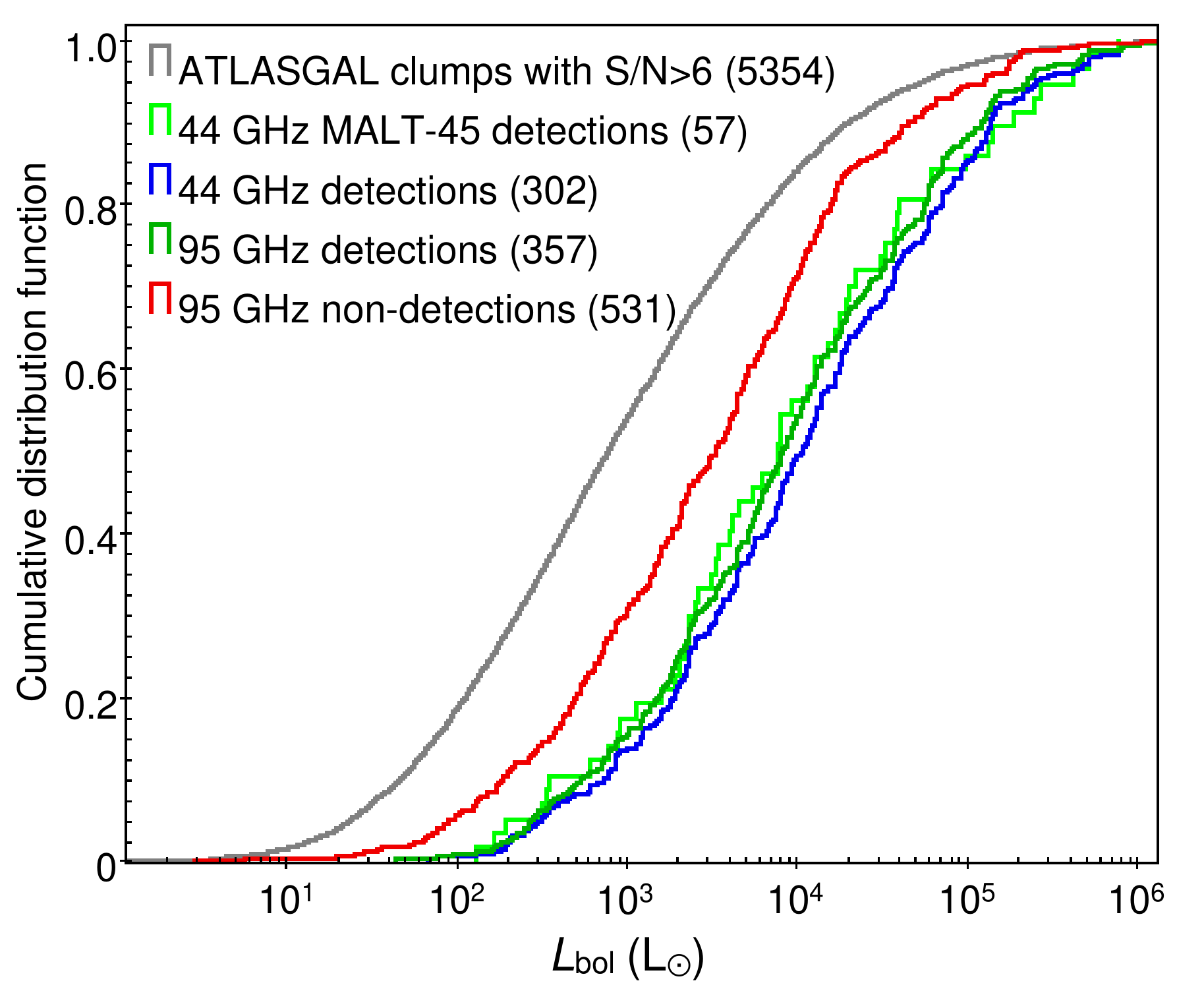}{0.40\textwidth}{(D)}}
\gridline{\fig{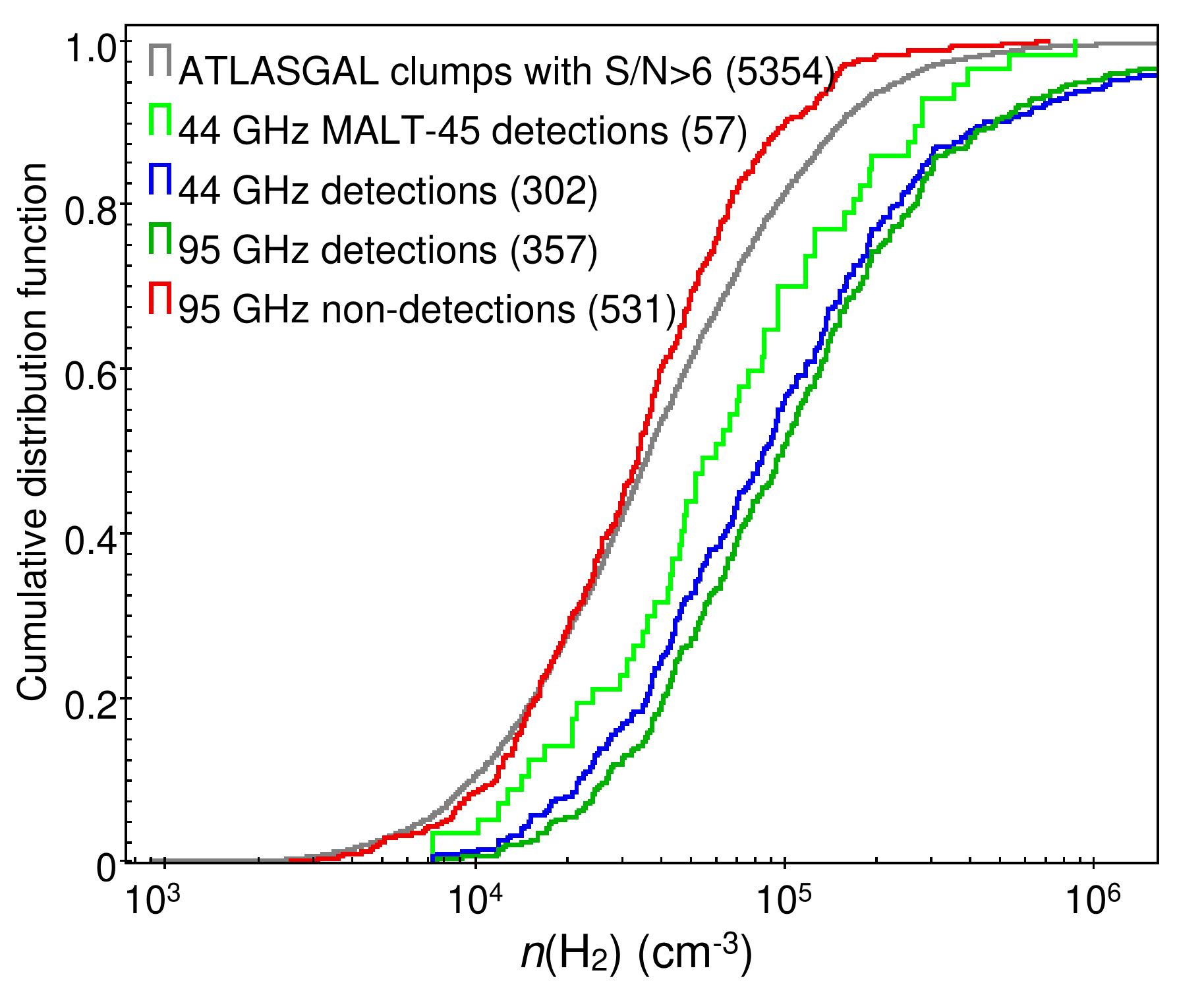}{0.40\textwidth}{(E)}
\fig{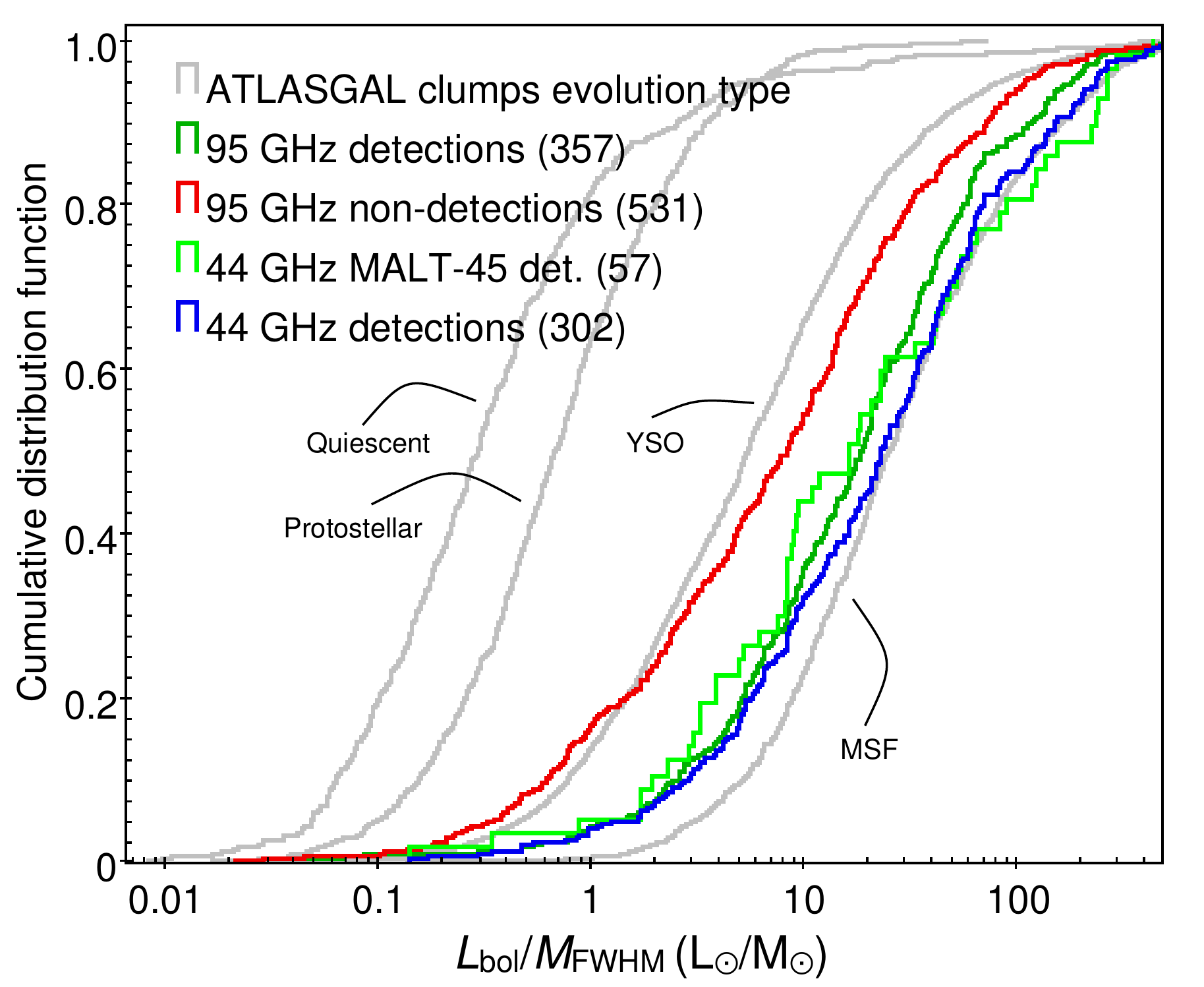}{0.4\textwidth}{(F)}}
\caption{Cumulative distribution plots of clump physical parameters ((A) peak flux density, (B) dust temperature, (C) FWHM mass, (D) bolometric luminosity, (E) number density and (F) luminosity to clump mass ratio) associated with class I methanol masers. The gray line shows all ATLASGAL sources (280\textdegree$<l<$60\textdegree, $|b|<1.5\degr$) with defined parameters. The green and red lines show those ATLASGAL clumps with an observed detection and non-detection of 95 GHz masers, respectively. The blue line is maser-associated clumps at 44 GHz. The light green line shows the sources with 44 GHz masers detected in the MALT-45 survey (\citealt{JOR15}; $330<l<335$, $|b|<0.5\degr$). In the last panel (F) grey lines show the different evolution types of all ATLASGAL sources, marked with labels.
\label{fig_hist}}
\end{figure*}

Figure~\ref{fig_hist} present{s} the cumulative distribution{s} of the clump physical parameters ($F_\mathrm{peak}$, $T_\mathrm{dust}$, $M_\mathrm{FWHM}$, $L_\mathrm{bol}$, $n\mathrm{(H_2)}$ and $L_\mathrm{bol}/M_\mathrm{FWHM}$) for sources both with associated class I methanol masers and those that are devoid of the maser emission. From the analysis of these plots, we conclude that clumps associated with class I methanol masers have some preferred regions of parameter space.

We have identified the clump parameters that have an association with 95 GHz masers by fitting Gaussian  profiles to the non-cumulative histograms of clumps physical parameters. Results of the fit presented in the Table~\ref{tab_fitpar}. From analysis of the fits we conclude that the physical parameters of maser-associated clumps peak at larger values compared to the sample of clumps with maser non-detections as well as the whole sample of ATLASGAL clumps. This can be seen in Figure~\ref{fig_hist} with almost all parameters except the FWHM clump mass being systematically higher than for the unassociated clumps. We note that the sample size for the whole ATLASGAL clump catalogue is different for each parameter, as not all \replaced{clump has}{clumps have} defined values of each physical parameter. The minimum sample size is 5354 sources for $M_\mathrm{FWHM}$, $L_\mathrm{bol}/M_\mathrm{FWHM}$ and $n\mathrm{(H_2)}$ parameters. The decreased sample size associated with the inclusion of only sources with S/N ratio more than 6 for correct FWHM mass and radii estimation.

The distribution of peak flux densities of ATLASGAL sources (see panel A in Figure~\ref{fig_hist}) reveals a pronounced shift between clumps associated with a 95/44 GHz methanol maser and those with 95 GHz methanol maser observations with non-detections. Due to the lack of information about 44 GHz maser non-detections, we used only 95 GHz maser non-detections. The distribution of ATLASGAL clump peak flux density  for sources associated \added{with} 95~GHz masers peaks at $\log F_\mathrm{peak}=0.372 \pm 0.013$.  
In contrast, clumps associated with 95~GHz non-detections peak at $\log F_\mathrm{peak}=0.008 \pm 0.008$. 
44 GHz masers have almost identical peak values of flux densities as those detected at 95 GHz (see panel A in Figure~\ref{fig_hist}). A Kolmogorov-Smirnov (KS) test of the peak flux densities of sources with and without detected 95 GHz masers gives a $p$-value of  $1.4\times10^{-15}$, meaning that the distributions can be considered to be significantly different.  

We also found pronounced shift\added{s} in bolometric luminosity, number density and luminosity to FWHM clump mass ratio for clumps associated with 95 GHz masers in comparison with \added{the} non-detection\added{s} of 95 GHz maser. A KS-test gives the $p$-values of $2.9\times10^{-8}$,  $6.6\times10^{-16}$ and $4.2\times10^{-10}$ for luminosity, number density and luminosity to FWHM clump mass ratio, respectively. We conclude that the differences in these parameters for detected and non-detected masers are statistically significant at $3\sigma$, i.e.  $p<0.0013$. The KS-test for \added{the} dust temperature gives \replaced{the}{a} $p$-value of 0.004, thus \added{the} difference is statistically significant at \added{a} 2$\sigma$ level ($p<0.005$).  In contrast,  there is no significant difference in the FWHM clump mass  for the different subsamples of maser-associated and non-detected clumps. \added{The} KS-test gives a $p$-value of 0.15, thus the null hypothesis about \added{the} same distribution cannot be rejected at \added{a} 2$\sigma$ level ($p>0.005$).

\begin{figure*}
\gridline{\fig{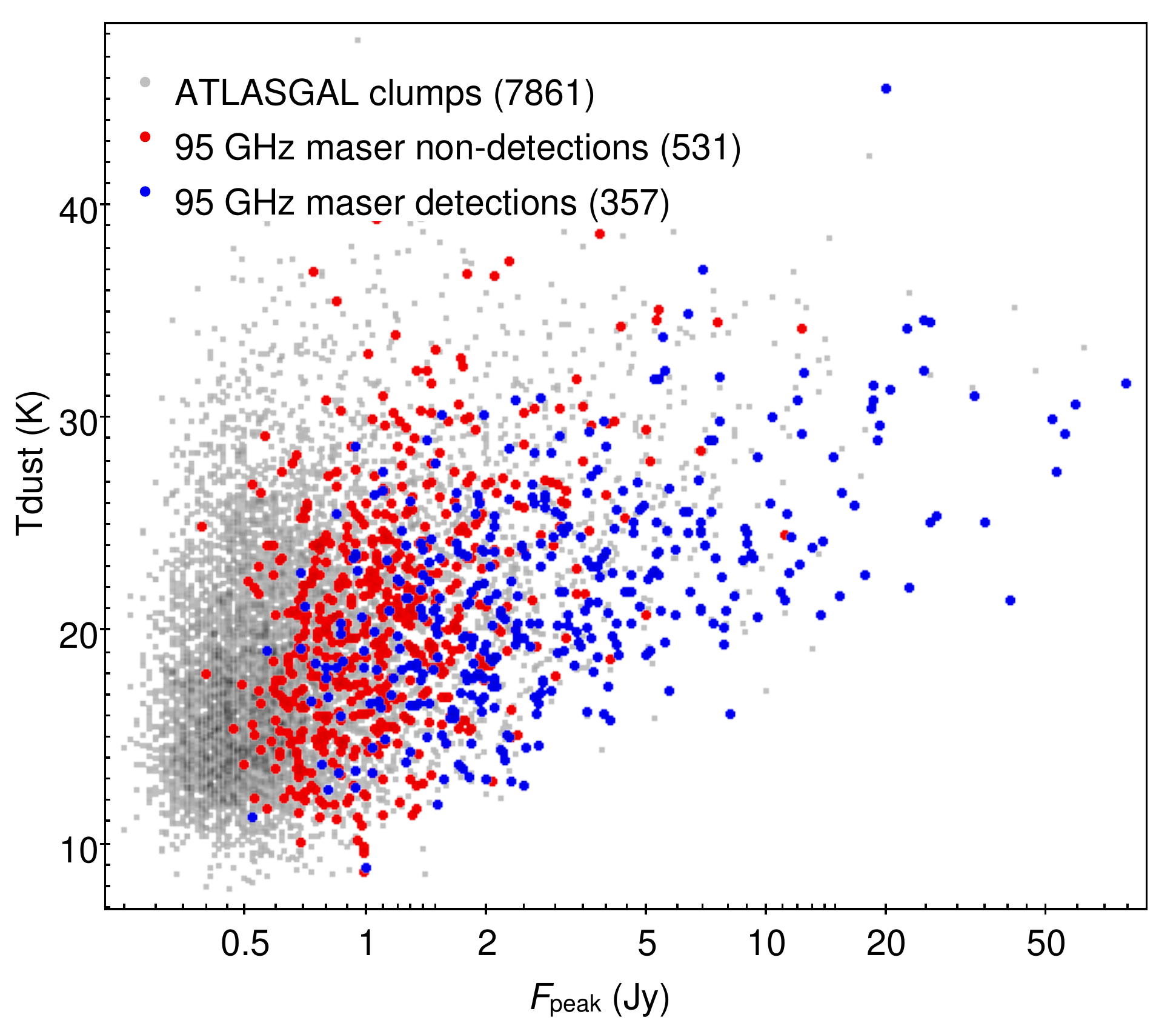}{0.45\textwidth}{(a)}
\fig{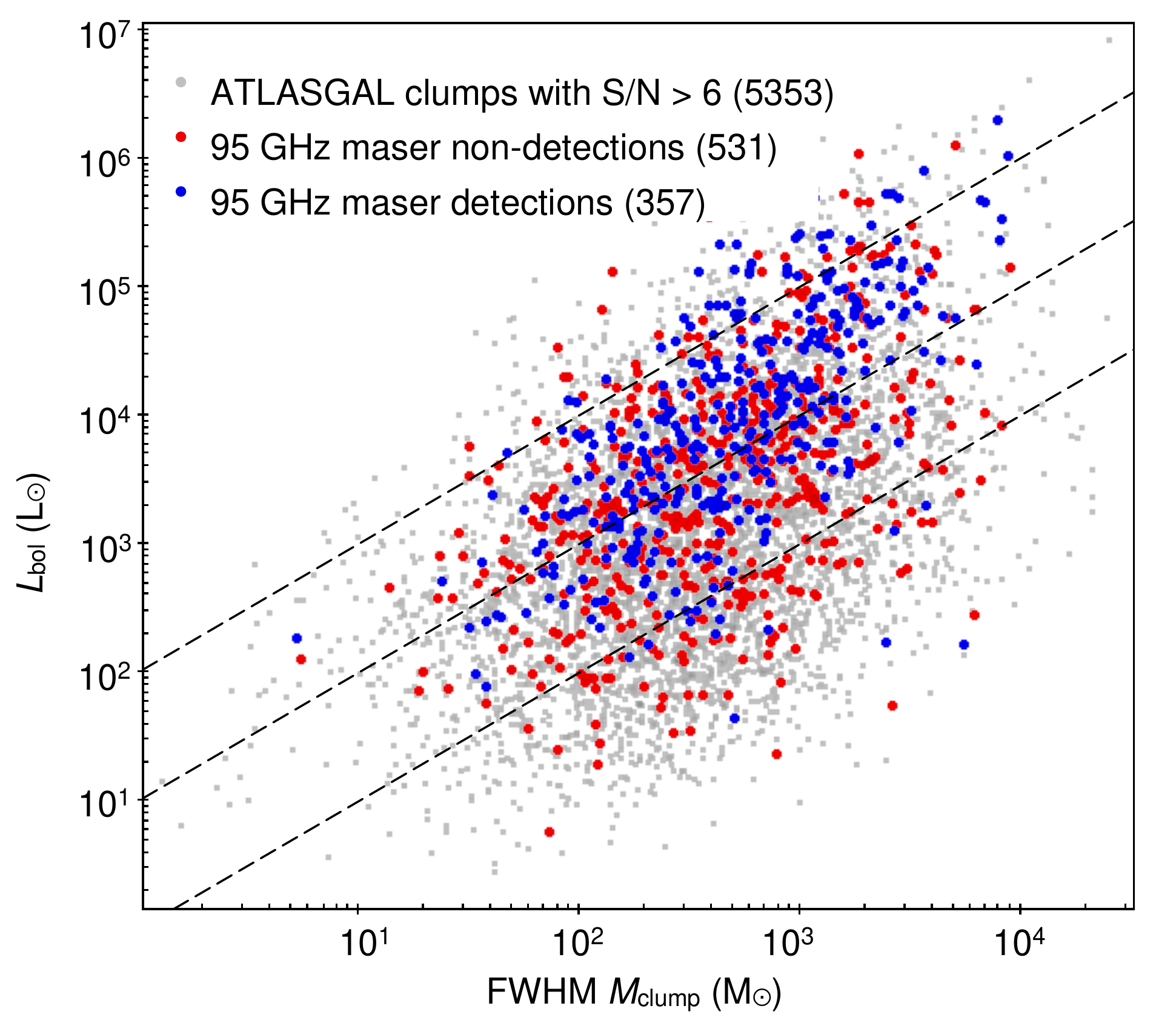}{0.45\textwidth}{(b)}
 }
\caption{
Distribution of ATLASGAL clumps peak flux density, dust temperature, luminosity and FWHM mass for all ATLASGAL clumps (gray dots), clumps associated with 95 GHz masers (blue dots), and clumps with non-detection of 95 GHz masers (red dots). Lower, middle and upper diagonal lines indicate the $L_\mathrm{bol}/M_\mathrm{clump}=1, 10$ and $100$~$\mathrm{L_{\odot}}/\mathrm{M_{\odot}}$, respectively.
\label{fig_phys_param}}
\end{figure*}

Analysis of the distribution of peak flux and dust temperatures (see left panel of Figure~\ref{fig_phys_param}) shows that 95 GHz methanol masers tend to arise in warmer sources, with a broad range of dust temperatures  ($12\lesssim T\lesssim 30$ (K) with no clearly preferred temperature within that range). The   distribution of clump mass and bolometric luminosity (right panel of Figure~\ref{fig_phys_param}) indicates that masers tend to arise in more luminous clumps not depending on its mass.

{ For 84\% of 95~GHz masers, the luminosity to mass ratio is between $1<L_\mathrm{bol}/M_\mathrm{FWHM}<100$~(M$_\mathrm{\odot}$/L$_\mathrm{\odot}$). We analyze the cumulative distribution function of luminosity to FWHM clump mass ratio (see panel F in Figure~\ref{fig_hist}) that is  considered to be a good diagnostic of the state of evolution of star-forming clumps (e.g. \citealt{molinari2008}). We additionally plot four evolution types of all ATLASGAL clumps from the classification of \cite{Urquhart18} in panel F of Figure~\ref{fig_hist}. It was found that ATLASGAL clumps associated with masers at 44 and 95 GHz have luminosity to mass ratio values between MSF (Massive Star Formation) clumps and YSO (Young Stellar Objects) clumps sub-samples, mainly associated with MSF clumps (as identified by the RMS survey; \citealt{lumsden2013}). These two samples are both mid-IR bright, i.e. associated with a 21-24 $\micron{}$ point source with a flux $>$ 2.6 mJy  \citep{Urquhart18}. The difference between these two samples is the presence of massive star-formation tracer\added{s} in MSF clumps, i.e. radio bright \hii{} regions, massive young stellar objects and methanol masers \citep{Urquhart14}.  Given that a fraction of masers (23\%, 84 out of 357) are associated with  the YSO clump sub-sample without massive star formation tracer (MSF sub-sample), we conclude that class I masers are formed prior to other tracers of star formation and provide the first evidence of star formation activity. This is consistent with the maser evolutionary diagram in star-formation regions \citep{Ellingsen07,Breen10}, where class I masers are considered to be the earliest  tracer of star formation.}

However, the preferred regions in parameter space may reflect a bias in the targeted observations of class\,I masers. To check this, we consider several samples of methanol maser\deleted{s} detections (see Figure \ref{fig_hist}). The first one is chosen to cover all 44 and 95 GHz maser observations in the ATLASGAL survey (280\textdegree$<l<$60\textdegree, $|b|<1.5\degr$). The second region ($330\degr<l<335\degr$, $|b|<0.5\degr$) is limited to the blind survey of class I methanol masers -- MALT-45 \citep{JOR15,JOR17}. The analysis of both distributions leads to the conclusion that both samples give the same results, but a larger sample has a smoother distribution. {The KS-test for a sample of 44 GHz masers detected in MALT-45 survey and the full sample of 44 GHz masers reveals that there are no significant differences between all considered parameters of the two maser distributions.
}

{The detection rate for the 44, 84 and 36 GHz masers remains unknown due to small samples of non-detected sources (92, 11 and 36, respectively). However, we can estimate the detection rate of 44 GHz masers using the sample of 307 sources that were observed at both 44 and 95 GHz. 271 sources from that sample were detected at 95 GHz, and 253 were detected at 44 GHz. Thus, the detection rates are comparable for both frequencies, and we assume that the sample of 44 GHz non-detected sources are similar to those non-detected at 95 GHz.} 

From analysis of the physical parameters of clumps associated with MMI at 95 and 44 GHz, we conclude that masers tend to arise in more luminous clumps with higher densities (both column density and number density) and temperatures  compared to the full sample of the ATLASGAL clumps. This is likely because protostars are associated with warmer and higher luminosity clumps and these are the sources that produce the shocks conducive to class I methanol maser emission. Furthermore, clumps associated with masers may be more heavily influenced by shocks than those devoid of masers and, hence, their average densities, luminosities and temperatures are all higher.

Shock wave propagation in clumps with higher densities and temperatures has a higher potential to increase densities and temperatures to the  levels required  for efficient class I methanol maser pumping. Clumps with higher luminosities have higher column densities, increasing the probability of acquiring high methanol column density in the shocked region, which is necessary to produce bright masers. The lower detection rates in very luminous and hot objects is likely because these objects have already evolved into \hii{} regions and have already started to disrupt their environments. However, class I methanol masers may arise even further while \hii{} region begin to expand and create the density waves with appropriate conditions (mainly density and temperature).

\begin{figure*}
\gridline{
    \fig{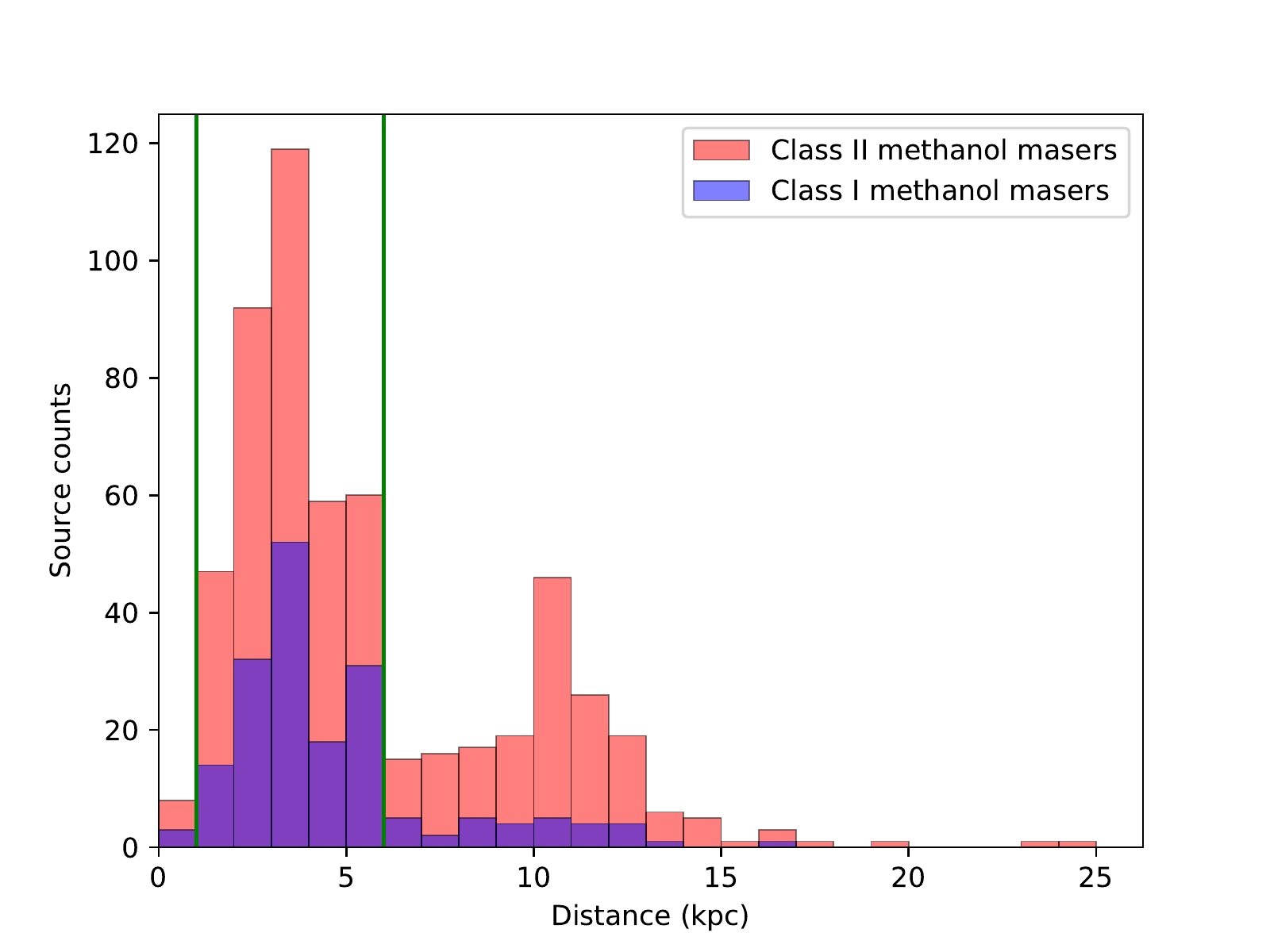}{0.5\textwidth}{(A)} 
    \fig{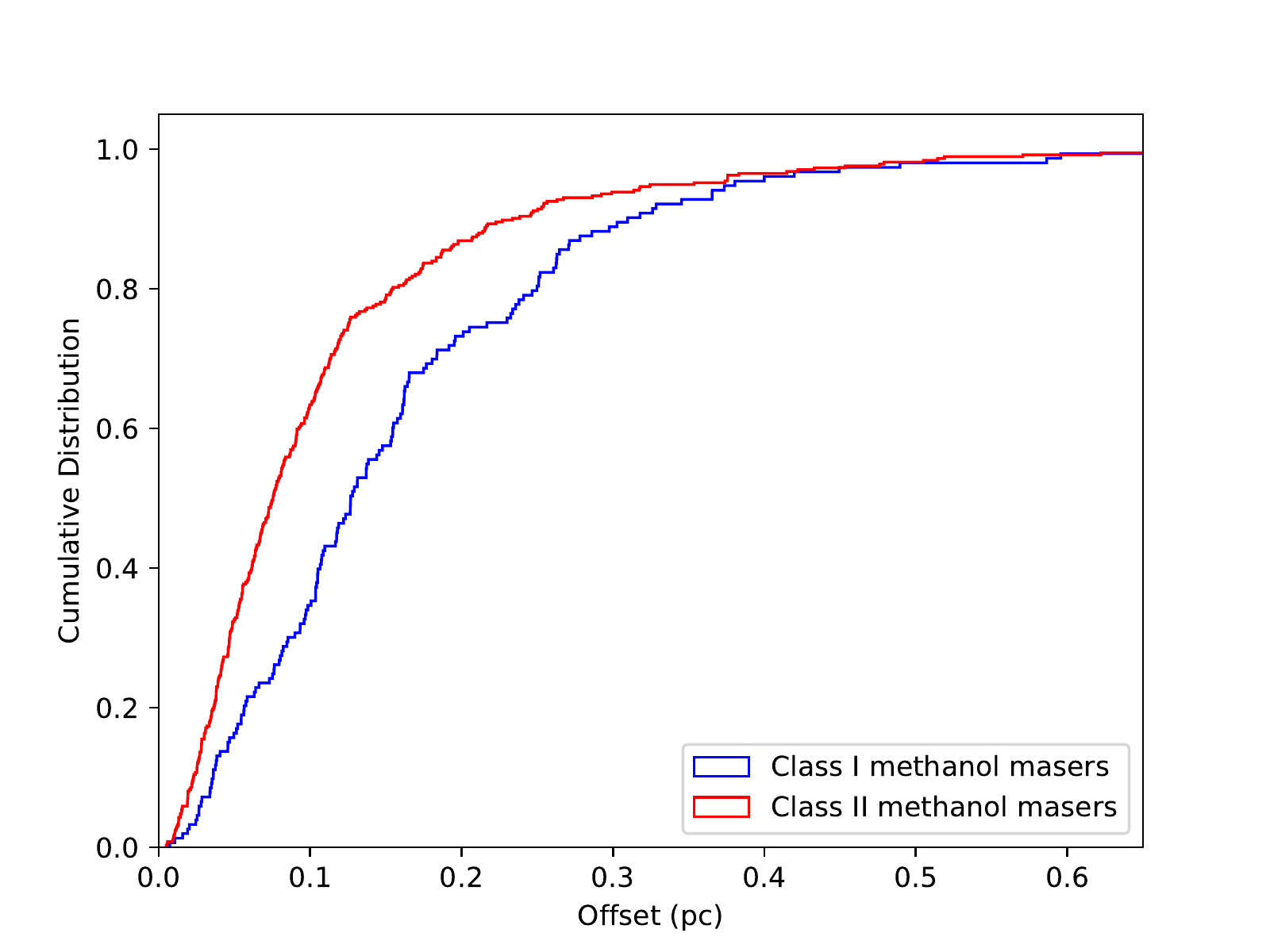}{0.5\textwidth}{(B)}
}
\caption{(A) Distribution of the distances to detected class I/II methanol masers. The green vertical line marks the distance-limited sample used for the construction of the cumulative offsets plot. (B) Cumulative distribution of  the physical offsets between ATLASGAL clump peak and methanol maser positions from the interferometric observations of a distance-limited sample ($1<D<6$~kpc). 
\label{fig_offset}}
\end{figure*}

\subsection{Separation between methanol masers and ATLASGAL clumps} \label{sec_offsets}

We have limited our analysis of class I and II methanol maser offsets from ATLASGAL clump peaks to the sample of masers with interferometric positions as listed in the maser database \citep{Ladeyschikov2019}. In cases where many maser spots are detected in a particular observation, we consider the brightest maser spot as the position for that group. In cases where several interferometric observations are available for a source, we select the median mean value of the available positions.  The physical distances to the sources were taken from \citet{Urquhart18}. We use a distance-limited sample in a range of 1-6~kpc to avoid any distance bias (see panel A in \deleted{the} Figure~\ref{fig_offset}). The positions of the brightest maser spots were matched with the ATLASGAL catalog using 30 arcsec matching radius, resulting in 153 matches for class I and 374 matches for class II methanol masers for distance-limited sample.  
The results are presented in the  panel B of Figure~\ref{fig_offset} in the form of the cumulative distribution of offsets between ATLASGAL clump peak and position of class I/II methanol masers. It shows the physical separation between ATLASGAL clump peak position and interferometric positions of class I/II methanol masers.

The collisional-radiative pumping of class I methanol masers should impact the linear separation between host dust clumps and positions of these masers. In contrast, the class II methanol masers at 6.7 GHz are pumped by infrared radiation and reside in the  circumstellar disks and inner parts of the outflows of high-mass YSOs,  which have sizes less than 1000 a.u. \citep{Sanna15,Sanna17}. \added{The} characteristic scale\deleted{s} of \added{the} class I methanol maser distribution is about 50 times larger \citep[e.g.][]{Voronkov14}.  Thus positions of class II masers can be regarded as the YSO position.  The right panel of Figure~\ref{fig_offset} clearly shows that there is a physical separation between ATLASGAL clumps and class I/II methanol masers, with class I masers being located at larger distances from the ATLASGAL clumps than the class II masers. This is in accord with the current understanding of the methanol maser origin and results of the previous analysis given in \citet[e.g.][]{Voronkov14}. while the class II masers are radiatively pumped in the accreting disk, but the class I are pumped by shocks and can be located further from the embedded protostar The mean distances between ATLASGAL clumps and methanol masers are found to be 0.16 and 0.11 pc for class I and II, respectively.

A KS-test comparing the offsets of the two methanol maser classes located between 1 and 6 kpc gives a $p$-value $\ll$0.0013, confirming that the distributions of class I/II masers within a clump are statistically significantly different.

\subsection{Correlation between MMI fluxes and ATLASGAL clump parameters}

{
\begin{table}
\caption{Partial correlation coefficients ($r$) between 95 GHz maser luminosities and ATLASGAL clumps physical parameters. Values in the brackets are the 95\% correlation confidence intervals. $N$ is the sample size, and $p$-value is the significance of the correlations. ``Limited'' refers to the distance-limited sample of ATLASGAL clumps within $1<D<6$~kpc. 
\label{tab_cc}}
\setlength{\tabcolsep}{2pt}
\small
 \begin{tabular*}{0.48\textwidth}{llccll}
\hline
    Parameter & Sample & $r$ & $N$ & $p$-value \\ 
  \hline
$\log[T_\mathrm{dust}]$ & Full & 0.19  [0.09, 0.29] & 357 & 0.027 \\ 
                        & Limited & 0.24  [0.12, 0.35] & 265 & $\ll 0.001$ \\ 
$\log[L_\mathrm{bol}]$  & Full &  0.40  [0.31, 0.48] & 357 & $\ll 0.001$ \\ 
                        & Limited & 0.39  [0.28, 0.49] & 265 & $\ll 0.001$ \\ 
$\log[M_\mathrm{FWHM}]$ & Full & 0.46  [0.37, 0.53] & 357 & $\ll 0.001$ \\ 
                        & Limited & 0.37  [0.26, 0.47] & 265 & $\ll 0.001$ \\ 
$\log[N\mathrm{(H_2)}]$ & Full & 0.28  [0.18, 0.37] & 357 & $\ll 0.001$ \\ 
                            & Limited & 0.41  [0.30, 0.50] & 265 & $\ll 0.001$ \\ 
$\log[L/M_\mathrm{FWHM}]$   & Full & 0.22  [0.12, 0.31] & 357 &  $\ll 0.001$ \\ 
                            & Limited & 0.28  [0.16, 0.39] & 265 & $\ll 0.001$ \\ 
$\log[n\mathrm{(H_2)}]$ & Full & 0.12  [0.02, 0.22] & 357 & 0.02 \\ 
                        & Limited & 0.31  [0.20, 0.42] & 265 & $\ll 0.001$ \\ 
      \hline
  {\smallskip}
\end{tabular*}
\end{table}
}

{We investigated the correlation between maser luminosity and ATLASGAL clump physical parameters by utilizing the non-parametric partial Spearman's correlation coefficients that removes the mutual dependence on the distance. We derive Spearman's correlation coefficients for each of the groups in two different samples -- full and distance-limited ($1<D<6$~kpc), as the maser flux for distant objects is limited by the telescope sensitivity, thus maser luminosity (as well as clump physical size) have higher uncertainty. The maser isotropic luminosity (hereafter luminosity) is calculated using $L_\mathrm{maser}=4\pi D^2 F_\mathrm{maser}$, where $D$ is the distance from \citet{Urquhart18}. The results of the correlation analysis are presented in \deleted{the} Table~\ref{tab_cc}. } 

{Analysis of the correlation coefficients has revealed that the maximum correlation in a full sample of maser-associated clumps ($N=357$) is found between the maser luminosity and the FWHM clump mass ($r=0.46$), as well as bolometric luminosity ($r=0.4$). However, the number density has a low correlation with maser luminosity in a full sample of maser-associated clumps. That can be due to uncertainty in mass and radius estimation for a distant objects. In the distance-limited sample ($1<D<6$~kpc), number density along with other parameters reveal almost similar correlations with $r\simeq0.3-0.4$. We conclude that more luminous, dense and massive clumps tend to produce more luminous masers.} 

\begin{figure}
\includegraphics[width=0.49\textwidth]{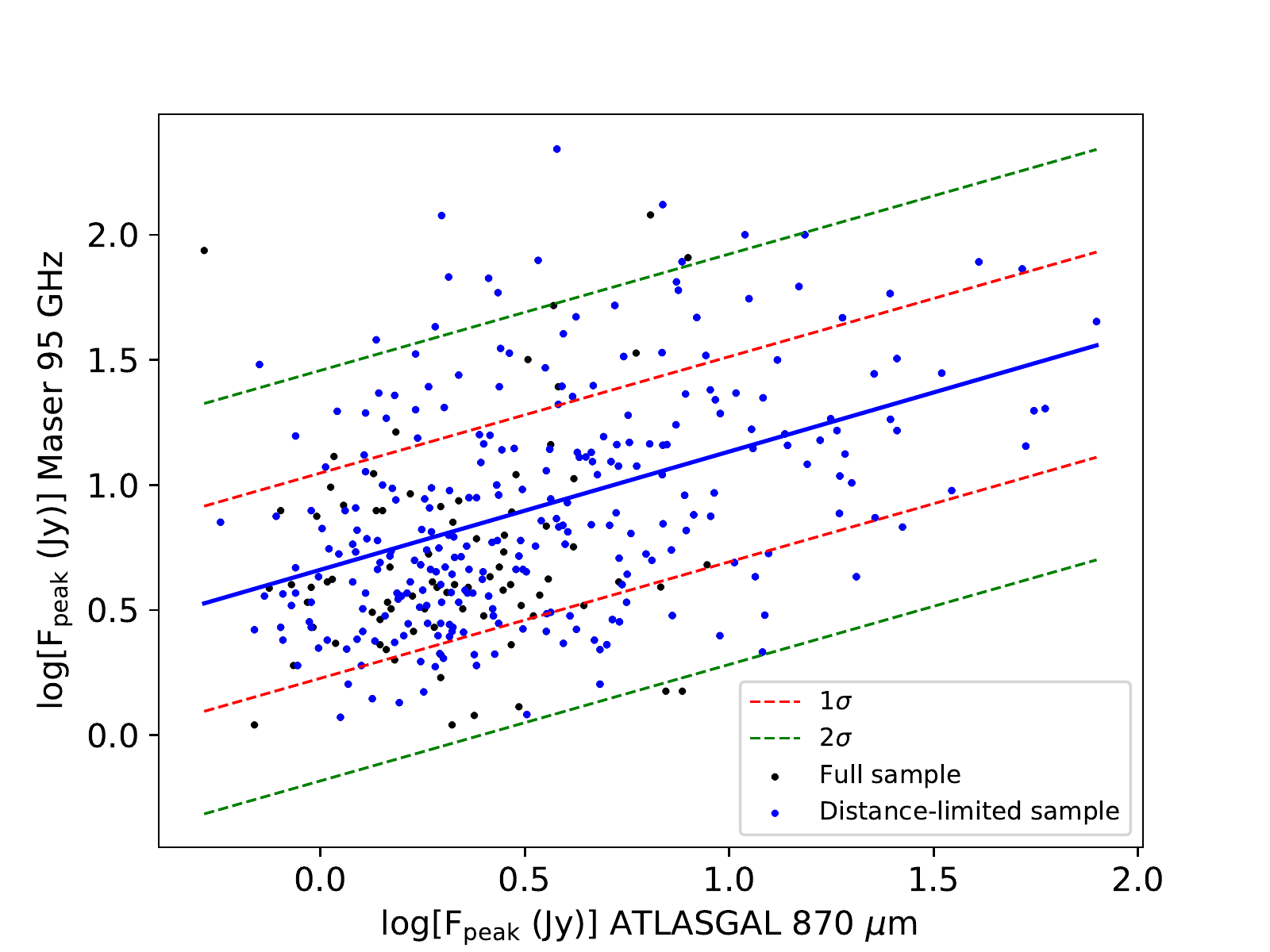} 
\caption{Distribution of ATLASGAL peak flux density against 95 GHz masers peak flux density. Black points and line refer to a full sample of maser-associated clumps, blue points and line refer to a distance-limited sample ($1<D<6$~kpc). The solid blue line is the linear fit (power-law in a log-scale) to the distance-limited sample. Red and green dashed line are $1\sigma$ and $2\sigma$ level of the deviation around the fit.  
\label{fig_corr}}
\end{figure}

{In \deleted{the} Figure~\ref{fig_corr} we present the log-log plot of ATLASGAL 870 \micron{} flux density against 95 GHz maser peak flux density. The correlation analysis for these two parameters gives the correlation coefficient $r=0.4$, similar to other parameters (see Table~\ref{tab_cc}). From a linear fit in log-log space we obtain the following} power-law relation $F_\mathrm{maser}=F_\mathrm{ATLASGAL}^{0.472\pm0.004}\times10^{0.661\pm0.003}$, where  $F_\mathrm{maser}$ and $F_\mathrm{ATLASGAL}$ both have units of Jy. The scatter in the data is quite significant ($\sigma[\log F_\mathrm{maser}]=0.41$) and so we are unable to assign a particular value of maser flux density from ATLASGAL peak flux density. However, in a distance-limited sample we find no masers below the $2\sigma$ level that have peak flux densities lower than predicted by the ATLASGAL peak flux density. Thus the empirical equation for the lower estimate of 95 GHz maser peak flux density can be written as:

{\begin{equation}
F_\mathrm{95GHz,lower}=F_\mathrm{ATLASGAL}^{0.472}\times10^{0.661-2\sigma}
\label{eq_maser_flux}
\end{equation}

\noindent where $\sigma=0.41$, $F_\mathrm{95GHz}$ and $F_\mathrm{ATLASGAL}$ are peak flux densities (in Jy) of 95 GHz masers and ATLASGAL sources, respectively. This equation may be used to estimate the minimum required noise level to detect 95 GHz methanol maser in a particular ATLASGAL source. 100\% of the masers in a distant-limited sample and 99\% of masers in a full sample have 95 GHz peak flux densities higher than the value determined from  Eqn.\,\ref{eq_maser_flux}}.

\subsection{Using ATLASGAL to search for MMI}\label{sec_survey}

We investigate the potential to use a subset of the ATLASGAL compact source catalog as a target list for further searches of class I methanol masers. As previously mentioned in Sect.\,\ref{sect_complet}, the MALT-45 blind survey reveals an MMI association rate with ATLASGAL sources of 95\% and the association of $\sim10$\% of ATLASGAL sources with MMI. The MALT-45 survey has a median noise level of 0.90 $\pm$ 0.09 Jy for 44 GHz class I methanol masers. 

The total number of ATLASGAL sources is much larger than the number of detected MMI. To use the ATLASGAL compact source catalog  efficiently for targeted observations, we need to develop a set of robust selection criteria  that will maximise the detection rate.  

In the first approach, we investigate the possibility of using the threshold on 870 \micron{} peak flux density of ATLASGAL sources for selecting the clumps with high maser detection rate. As was shown in \deleted{the} Table~\ref{tab_fitpar}, there is a pronounced shift between flux \replaced{densities}{density} and H$_2$ column density of the detected and non-detected samples. Thus, the threshold on ATLASGAL source peak flux density or H$_2$ column density  may significantly increase the detection rate, but miss a fraction of low-brightness ATLASGAL clumps associated with masers. To avoid this and cover at least 96\% of detected masers associated with ATLASGAL clumps we have used either of the following thresholds:

\begin{equation}
F_\mathrm{peak}>0.85~\mathrm{Jy}
\label{eq_flux_thresh}
\end{equation}
\begin{equation}
\log{}N(\mathrm{H}_2)>22.3~\mathrm{cm}^{-2}
\label{eq_density_thresh}
\end{equation}

{The analysis of Eqn.\,\ref{eq_flux_thresh} lead to the conclusion that the 95 GHz maser detection rate toward ATLASGAL sources satisfying that criteria is $\sim50\%$ for an available sample of 784 objects. The typical $\sigma$ level for achieving this detection rate is 0.2 Jy. The same analysis for Eqn.\,\ref{eq_density_thresh} lead to the detection rate of $\sim45\%$ for the sample of 793 objects.}

{A similar analysis of the BGPS survey with sources satisfying   $\log(S_\mathrm{int})\leq-38.0+1.72\log(N_\mathrm{H_2}^\mathrm{beam})$ and $\log(N_\mathrm{H_2}^\mathrm{beam}))\geq22.1$ \citep{CHE12} lead to the detection rate of 30\% with a sample of $\sim$1000 sources. This value is close to the observed detection rate of 29\% \citep{CHE12} with a sample of 214 sources.}

\section{Conclusions}

From the analysis of MMI and ATLASGAL sources, we have found a tight  physical correlation between them -- almost 100\% of class I methanol maser sources have ATLASGAL counterparts. Analysis of the physical parameters of masers-associated ATLASGAL clumps leads to the following conclusions:
\begin{itemize}

\item {The distribution of ATLASGAL clumps that have been the targets of methanol maser observations reveals a statistically significant differences in physical parameters between 95/44 GHz maser-associated clumps and ATLASGAL clumps without detection of 95 GHz masers. The following parameters have statistically significant difference: peak flux density,  number density,  column density, bolometric luminosity,  luminosity to clump mass ratio and dust temperature.} 

\item {Masers tend to arise in more luminous clumps with higher densities and slightly higher temperatures (significant at the 2$\sigma$ level) compared to the whole sample of the ATLASGAL clumps.   We conclude that the warmer and higher luminosity clumps host protostars and it is the shocks they generate that drive the masers. From the analysis of partial correlation coefficients between 95 GHz maser luminosities and ATLASGAL clump physical parameters, we conclude that more luminous, dense and massive clumps tend to produce more bright masers. }

\item {23\% of 95 GHz masers (84 out of 357) are associated with ATLASGAL clumps that  are mid-IR bright, i.e. associated with a 21-24 $\micron{}$ point source with a flux $>$ 2.6 mJy \citep{Urquhart18}, but display no other tracers of massive star formation, i.e. radio bright \hii{} regions, massive young stellar objects or class II methanol masers \citep{Urquhart14}.  This leads us to conclude that class I masers are formed prior to other tracers of star formation and provide the first evidence of star formation activity. This is consistent with the maser evolutionary diagram in star-formation regions \citep{Ellingsen07,Breen10}, where class I masers  are some of the earliest tracers of star formation. }

\item {The physical separation between the center of the clumps and the brightest maser spots is significantly smaller for class II masers compare to class I masers. This is consistent with our understanding of the pumping mechanisms and their connection with the star formation process-- the class II masers are radiatively pumped in the accreting disk, but the class I are pumped by shocks and can be located further from the embedded protostar. The mean distances between ATLASGAL clumps and methanol masers are found to be 0.16 and 0.11 pc for class I and II, respectively.}

\item {We investigate the potential of the ATLASGAL compact source catalog as a target list for the search of class I methanol masers. The threshold values of F$_\mathrm{ATLASGAL}>0.85$~Jy or $\log N(\mathrm{H_2})>22.3$~cm$^{-2}$ gives a coverage of 95\% known class I methanol masers with estimated detection rate of $\sim50$\% at 1$\sigma$ level of $\sim$0.2 Jy for maser observations. }

\end{itemize}

\section*{Acknowledgements}

The work of DAL in the Section 3 was supported by the Ministry of Education and Science of Russia (the basic part of the State assignment, RK no. FEUS-2020-0030). The work of DAL and AMS in the  Section 4.1 was supported by Russian Science Foundation grant 18-12-00193. The work of DAL in the Section 4.2-4.4 were supported by Russian Foundation for Basic Research through research project 18-32-00605.

\bibliography{bibtex}{}
\bibliographystyle{aasjournal}


\listofchanges

\end{document}